%% file: main.tex
\documentclass[letterpaper,twocolumn,10pt]{article}
\usepackage{usenix,epsfig,endnotes}
\usepackage{times}

\usepackage{datetime}
\usepackage{url}
\usepackage{breakurl}
\usepackage[breaklinks]{hyperref}
\usepackage{graphicx}
\usepackage{subfigure}
\usepackage{color}
\usepackage{multirow}
\usepackage{makecell}

\newcommand{\TODO}[1]{{\color{red} {\bf #1}}}
\newcommand{\tf}{{TensorFlow}}

\begin{document}

\date{}

\title{\Large \bf RPC Considered Harmful: \\ Fast Distributed Deep Learning on RDMA} 

\author{\rm Jilong Xue, Youshan Miao, Cheng Chen, Ming Wu,  Lintao Zhang, Lidong Zhou \\ 
	Microsoft Research}


\maketitle

\input{abstract}

\input{intro}
\input{background}

\input{design}

\input{imp}
\input{eval}

\input{related}

\input{conclusion}

\def\UrlBreaks{\do\/\do-}
\bibliographystyle{acm}
\bibliography{./bib/papers}

\end{document}

%% file: abstract.tex
\subsection*{Abstract}

Deep learning emerges as an important new resource-intensive workload and has been successfully applied in computer vision, speech, natural language processing, and so on. 
Distributed deep learning is becoming a necessity to cope with growing data and model sizes.
Its computation is typically characterized by a simple tensor data abstraction to model multi-dimensional matrices, a data-flow graph to model computation, and iterative executions with relatively frequent synchronizations, thereby making it substantially different from Map/Reduce style distributed big data computation.

RPC, commonly used as the communication primitive, has been adopted by popular deep learning frameworks such as TensorFlow, which uses gRPC. 
We show that RPC is sub-optimal for distributed deep learning computation, especially on an RDMA-capable network.
The tensor abstraction and data-flow graph, coupled with an RDMA network,
offers the opportunity to reduce the unnecessary overhead (e.g., memory copy) without sacrificing programmability and generality.
In particular, from a data access point of view,
a remote machine is abstracted just as a ``device'' on an RDMA channel, with a simple memory interface for allocating, reading, and writing memory regions.
Our graph analyzer looks at both the data flow graph and the tensors to optimize memory allocation and remote data access using this interface.
The result is up to $25\times$ speedup in representative deep learning benchmarks against the standard gRPC in TensorFlow and up to 169\% improvement even against an RPC implementation optimized for RDMA, leading to faster convergence in the training process.

%% file: intro.tex
\section{Introduction}
\label{sec:intro}

Deep learning, in the form of deep neural networks (DNN), is gaining popularity thanks to its huge success in areas such as speech, vision, and natural language processing. 
There is a trend of using deeper, more complex neural network models trained with increasingly larger data sets.
Such a model often takes hours, days, or even weeks to train on a CPU/GPU cluster.
The deep learning computation in training a model involves multiple iterations with rather frequent synchronizations.
The performance therefore often critically depends on the efficiency of cross-machine communication, including its ability to leverage emerging network technology, such as 
Remote Direct Memory Access (RDMA).

Remote Procedure Call (RPC) is a widely used general-purpose communication paradigm.
In addition to data transfer, RPC takes care of data serialization and deserialization for various data types, manages communication buffers, and handles message assembly and batching automatically.
Even with RDMA, RPC can be used to help mediate concurrent (remote) writes to the same data~\cite{fasst16}.
It is therefore natural for deep learning frameworks such as  \tf{}~\cite{tensorflow16} to adopt gRPC, a form of RPC, as its communication abstraction.

In this paper, we argue against using RPC for distributed deep learning computation, especially on an RDMA-capable network.
This is because (i) deep learning computation uses tensor (or multi-dimensional matrix) as the main data type, which consists of a plain byte array as tensor data and a simple schema as meta-data specifying the shape and element type of the tensor. A tensor is often of a sufficiently large size (tens of KB to MB) and its metadata/data sizes often static. Using RPC  for tensor data transfer does not provide evident advantage on programmability or efficiency; and 
(ii) using RPC typically involves memory copy to and from RPC-managed  communication buffers. Zero-copy cross-machine tensor transfer is possible with RDMA because the source and destination tensors can be appropriately allocated in the RDMA memory region and known statically.  
We therefore advocate a simple and almost trivial interface that exposes a remote machine as a ``device'' from a data access point of view. This ``device'' is connected through an RDMA-based channel that exposes control for parallelism.
Remote memory regions can be allocated and directly accessed through this ``device'' interface, much like the case for a local GPU.
This maps naturally to the underlying RDMA network that provides direct remote memory access.
It is worth pointing out that previous work on efficient communication on RDMA often uses RPC (e.g., for writes) partly because they are focusing on variable (and often small) size data transfer for key/value stores, where they can benefit from batching in RPC and from mediating concurrent remote writes to the same region through RPC~\cite{Jinyang13,farm14,herd14,rdma16}. Neither is necessary in our case.

We have designed a zero-copy cross-machine tensor transfer mechanism directly on our ``device'' interface. This is done through a combination of static analysis and dynamic tracing on the data-flow graph for the computation in order to (i) figure out whether the size of each tensor that needs to be transferred across server can be statically known at the compile time, (ii) assess whether such a tensor should be allocated statically (for better efficiency) or dynamically (for reduced memory footprint), (iii) ensure allocation of the tensors on both the sending and receiving ends in the RDMA memory regions, (iv) identify the source and destination addresses of tensors for RDMA-based transfer.

We have implemented an efficient RDMA-based ``device'' library, and integrated it with our graph analysis and tracing mechanism into the data-flow graph runtime of \tf{}~\cite{tensorflow16} for tensor data transfer in distributed deep learning computation.
The experiments show that our proposed techniques help \tf{} achieve up to $25\times$ speedup in representative deep learning benchmarks compared with its original gRPC-based communication mechanism and even up to 169\% improvement against an RPC implementation optimized for RDMA.

%% file: background.tex
\section{Background and Problems}
\label{sec:bg}

\subsection{Deep Learning Data-Flow Graph}

\begin{figure}[t]
	\centering
	\includegraphics[width=.95\linewidth]{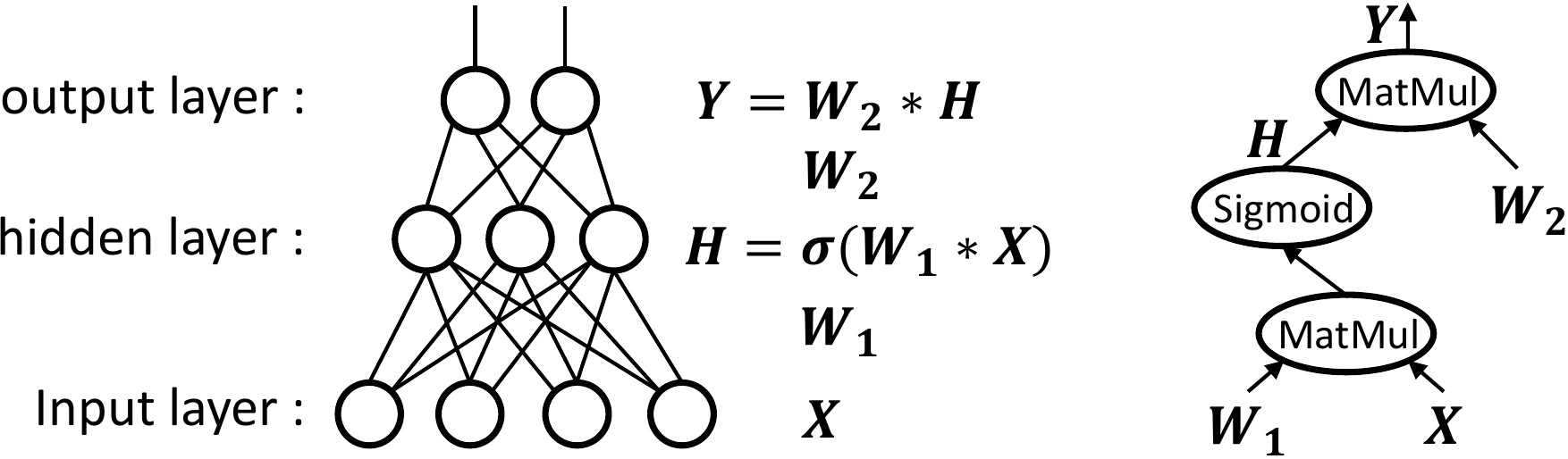}
	\vskip -2ex
	\caption{Example of a vanilla neural network (left) and the data-flow graph (right) of its forward computation. $\sigma$ is the non-linear \emph{Sigmoid} function. Bold symbols are the variables representing tensors.}
	\vskip -1ex
	\label{fig:nn}
\end{figure}

Deep neural network describes a layered machine learning model that consists of neurons connected through synapses. The layered structure enables the model to learn hierarchical features from the raw input data. Each layer normally represents a linear transformation on its inputs followed by some non-linear activations. The left side of Figure~\ref{fig:nn} shows an example of a vanilla deep neural network. The parameters to learn in this model are the weights of the connections between neurons of different layers. The computation on this model can be naturally expressed using a data-flow graph where the nodes represent the computations at layers and the edges represent the data flowing between the dependent nodes. In deep learning scenarios, the major data type flowing in the graph are tensors (i.e., multi-dimensional matrices) because most deep learning algorithms are expressed as mathematical models on matrices. The right side of Figure~\ref{fig:nn} shows an example data-flow graph expressing a forward computation on the neural network in the figure from the raw input to upper layers. Through supporting this data-flow graph representation for deep learning computation, frameworks~\cite{torch11,cntk14,mxnet15,theano16,tensorflow16} can allow developers to conveniently implement variant forms of neural networks that can be very complex.

\begin{figure}[t]
	\centering
	\includegraphics[width=0.65\linewidth]{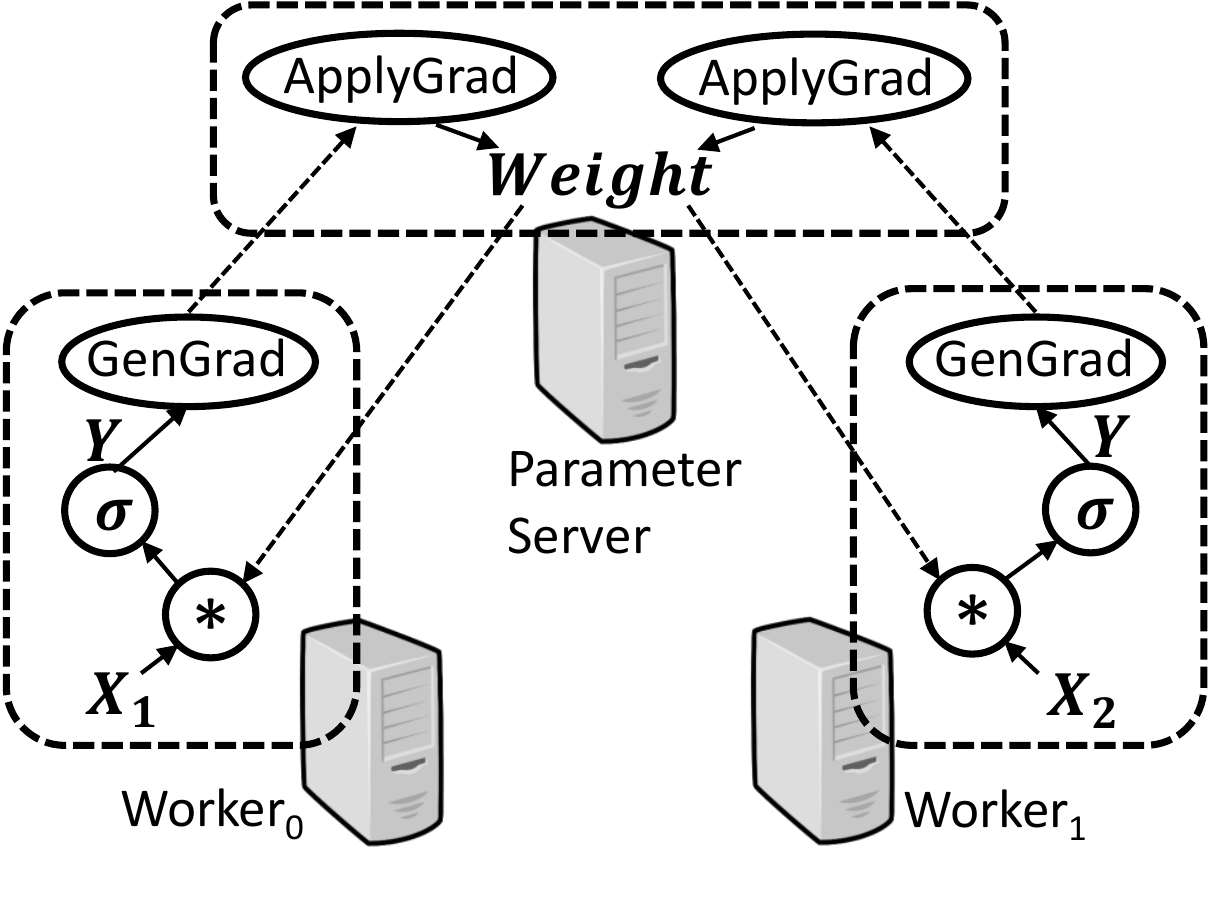}
	\vskip -2ex
	\caption{Example of distributed data-flow computation of deep learning with data-parallelism in parameter server architecture. Dotted-line arrow refers to the cross-server data flow. For simplicity and clarity, \emph{GenGrad} and \emph{ApplyGrad} represent the sub-graphs of computing and applying gradients, respectively. \emph{Weight} is the tensor representing the model parameters which is shared by the graph replicas.}
	\vskip -2ex
	\label{fig:dist}
\end{figure}

The training process of a deep neural network can be very time-consuming because it is hard to learn complex hierarchical features and consequently the computation often requires processing very large training data. In order to scale out the computation, distributed deep learning can be applied by replicating and partitioning the data-flow graph onto multiple servers to execute in data-parallel or model-parallel fashion. The data flow between the graph nodes across partitions will be fulfilled through the underlying communication layer during the computation. Figure~\ref{fig:dist} shows an example of distributed deep learning computation in data-parallelism. A data-flow graph is replicated on two workers and each replica is partitioned among a worker and a parameter server. Employing such distributed data-flow graph brings much convenience and flexibility on playing model-parallelism which is critical when the deep learning model size is large.


The deep learning process involves iterative executions over the data-flow graph for multiple mini-batches of training data. Therefore, after the data-flow graph is created and before the computation starts, a deep learning framework can reasonably take some time to analyze the graph and optimize the execution. In the graph analysis phase, some useful information can be extracted and potentially leveraged by components in lower execution layer to improve runtime efficiency. One example of such information is the addresses of the tensor data that need to be transferred across servers. This information can be obtained statically sometimes, because the shapes of some tensors do not change during the entire computation; e.g., in the case of the tensors representing the model parameters. The framework can then arrange the placement of those tensors in memory before the execution of the data-flow graph. It is therefore feasible to design an appropriate abstraction for the communication layer to accept such information to improve its efficiency.

\subsection{RPC Abstraction}
\label{sec:rpc}
Remote Procedure Call (RPC)~\cite{rpc84} is a common abstraction for communication across servers. It allows users to implement a procedure that can be invoked remotely as if being invoked locally. With RPC, users only need to focus on the implementation of the functional logic of the remote procedure without caring about the underlying communication-related details. In addition, RPC is often used to pass structured messages since it usually integrates serialization and deserialization functionalities. There are many existing designs and (open-sourced) implementations of RPC~\cite{grpc,thrift,zmq,herd14} from industry and research communities. They have been extensively applied in many distributed systems~\cite{hadoop09,storm14,mxnet15,tensorflow16,fasst16}.



The RPC abstraction normally assumes that a communication channel between a pair of endpoints can be used to transfer arbitrary types of messages with
respect to data schema and size at any point during runtime. 
This convenience is not particularly beneficial in the deep learning scenario mainly because the major data abstraction is the \emph{tensor}, whose meta-data contains only simple schema with shape and element type information.
There is an inherent cost associated with providing this general convenience: 
It makes hard for the communication library to be aware, in advance, of which user buffer the received message should be directly delivered to. Therefore, a common way is to use a fixed in-library buffer to receive a message from the operating system layer and then copy the data to the appropriate user buffer. An in-library buffer is associated with each channel and should have a limited size; otherwise, there will be a scalability issue of memory consumption when the cluster of servers become large. And also, when the caller side wants to transmit a message larger than the receiver side buffer, the message has to be split into multiple fragments with each having some header information added for re-assembling at receiver side. This often enforces extra data copy at the sender side. These data copy overheads are proportional to the message size, and hence can be significant when message is large. Without re-designing the abstraction, it is hard, if not impossible, to eliminate these overheads completely in the communication layer.

\subsection{Remote Direct Memory Access}
Remote Direct Memory Access (RDMA) is an emerging fast network technology that allows one computer to  access directly the memory of a remote computer without involving the operating system at any host. 
With the technology maturing and cost competitive, RDMA has found its way into data centers and is gaining popularity~\cite{Jinyang13}. 

The user interface to issue RDMA operation is through functions called \emph{verbs}. There are two types of verbs semantics, memory verbs and messaging verbs. The memory verbs include one-sided RDMA reads, writes, and atomic operations. These verbs specify the remote memory address to operate on without involving the remote CPU. This lack of CPU overhead at remote side makes them attractive. The messaging verbs include the send and receive verbs, which involve the remote side CPU. Verbs are posted by applications to queues that are maintained inside the RDMA NIC. Queues always exist in pairs with a send queue and a receive queue forming a queue pair (QP). Each queue pair has an associated
completion queue (CQ), which the RDMA NIC fills in upon completion of verb execution. RDMA transports can be either reliable or unreliable, and either connected or unconnected (also called datagram). In our work, we always use reliable connected transport.  

RDMA networks provide high-bandwidth and low latency: NICs with 100 Gbps bandwidth and $\sim$2$\mu$s round-trip latency are commercially available. The high-bandwidth of RDMA and its kernel-bypassing nature make any communication related computation overhead significant. We observe that removing the extra copy of message data can evidently improve the communication efficiency. Simply building a general RPC abstraction over RDMA makes it hard to avoid these data copy overhead. For example, the message passing mechanism used in FaRM~\cite{farm14} RPC employs a fixed ring buffer with each channel on the receiver side and may suffer from the problem described in \S\ref{sec:rpc}.

The one-sided memory read/write semantic of RDMA allows a zero-copy communication across servers as long as the remote address is known. In the deep learning computation scenarios, the data-flow graph analysis can help arrange the in-memory placement of tensors and provide such information to the underlying communication layer. This leads us to believe that exposing a simple memory copy interface directly is the most appropriate  because tensor is the major data type that need to be transferred across servers during deep learning computation. 

%% file: design.tex
\section{Design}
\label{sec:design}
We now present our design for the RDMA device communication abstraction, the tensor transfer mechanisms, and the integration with the deep learning data-flow graph analysis.

\subsection{RDMA Device Abstraction}

\begin{figure}
	\centering
	\includegraphics[width=0.9\linewidth]{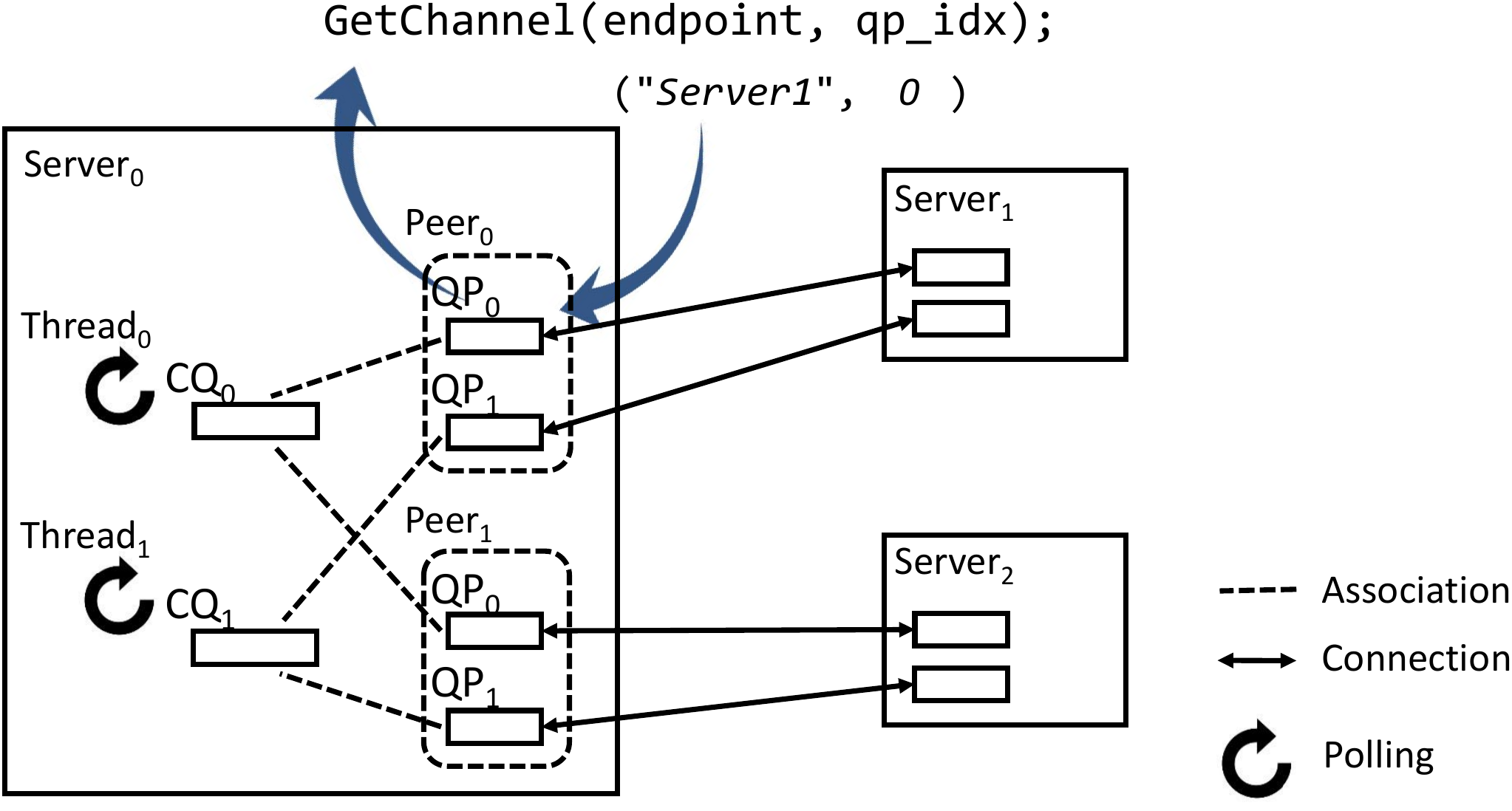}
	\vskip -2ex
	\caption{Overview of the architecture design of the RDMA memory copy library. QPs are created and grouped peer by peer and associated with CQs in a round-robin way.}
	\vskip -3ex
	\label{fig:netlib}
\end{figure}


Our communication library provides a simple abstraction for each RDMA NIC as an RDMA device (or \emph{device} for short when there is no confusion). The device provides an interface to allocate and free a memory region that can be accessed by other devices remotely. Given a remote device specified as an endpoint (i.e., IP address and port), users can acquire a channel from the local device object that connects the local device and the remote one. A channel corresponds to an RDMA QP and provides a memory copy interface for cross-server data transfer, which takes a local and a remote memory regions and a transfer direction as arguments. The actual data transfers are performed using the one-sided RDMA read/write verbs. To use the memory copy interface, one has to know the address of the to-be-accessed remote memory region. The library, therefore, also provides a simple vanilla RPC mechanism implemented using the RDMA send/recv verbs for this auxiliary purpose of distributing remote memory addresses. This address distribution process is often not on the critical path of the application, and hence not performance critical.

The RDMA device is configured with the number of CQs per device and the number of QPs for each connected peer remote device. The library maintains a thread pool with each thread polling a specific CQ for completion of RDMA events. When establishing a connection to a remote peer device, it evenly spreads the associations of the created QPs with the CQs in a round-robin fashion. The channel acquiring interface allows users to specify the specific QP that the channel uses. Through this interface, a multi-threaded workload, e.g., the deep learning graph execution runtime, is able to balance the loads and synchronization cost over the QPs and CQs to achieve good parallelism and communication efficiency. Figure~\ref{fig:netlib} shows an overview of this design.

\subsection{Transfer with Static Placement}
\label{sec:static}

\begin{figure}
	\centering
	\includegraphics[width=0.9\linewidth]{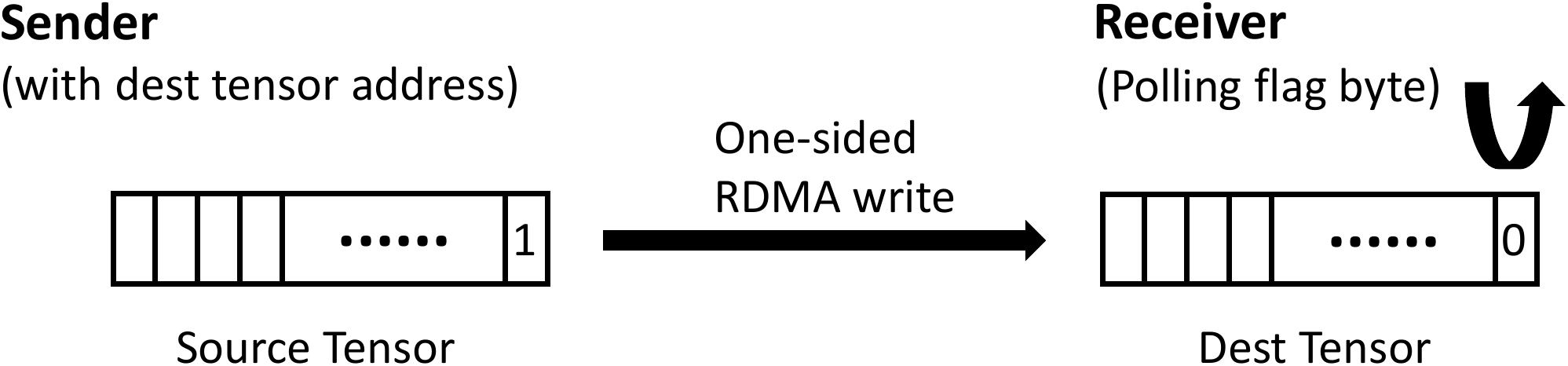}
	\vskip -2ex
	\caption{Transfer statically placed tensor through one-sided RDMA write.}
	\vskip -2ex
	\label{fig:staticmem}
\end{figure}

During the graph analysis phase, the shapes of some tensors can be statically decided and will not change during the entire computation.
Examples include those  tensors holding the parameters of the model to be trained. Given this information, the analysis engine can allocate the memory regions for these tensors beforehand and fix their placement during the computation. If the content of such a tensor relies on that of a remote one, its address, which is remotely accessible, is distributed to the server that holds the remote upstream tensor before the computation. The sender side of the tensor transfer can then use the memory copy interface to write the content of the downstream tensor at receiver side directly during the computation. 

The receiver side needs to know whether the content of the downstream tensor has been written in full. This is achieved through introducing a flag byte at the tail of the tensor memory region. The sender transfers the tensor content together with the flag byte set. The transfer is conducted from low address to the higher. The flag is then the last byte being transferred. Many RDMA NICs (including the ones we are using) guarantee that the RDMA writes are performed in an ascending address order (same as reported in FaRM~\cite{farm14}). So, once the flag byte is delivered, the entire tensor content must have been written in full. The receiver periodically polls the flag byte of the downstream tensor. Once the tensor transfer completes, it clears the flag for future use and then activates the graph nodes, which depend on this transferred tensor, to make them ready to execute. Figure~\ref{fig:staticmem} illustrates this mechanism. The receiver side polling may have lower priority than other ready tasks so does not block them and only introduces minor cost.

\subsection{Transfer with Dynamic Allocation}
\label{sec:dyn}

\begin{figure}
	\centering
	\includegraphics[width=0.9\linewidth]{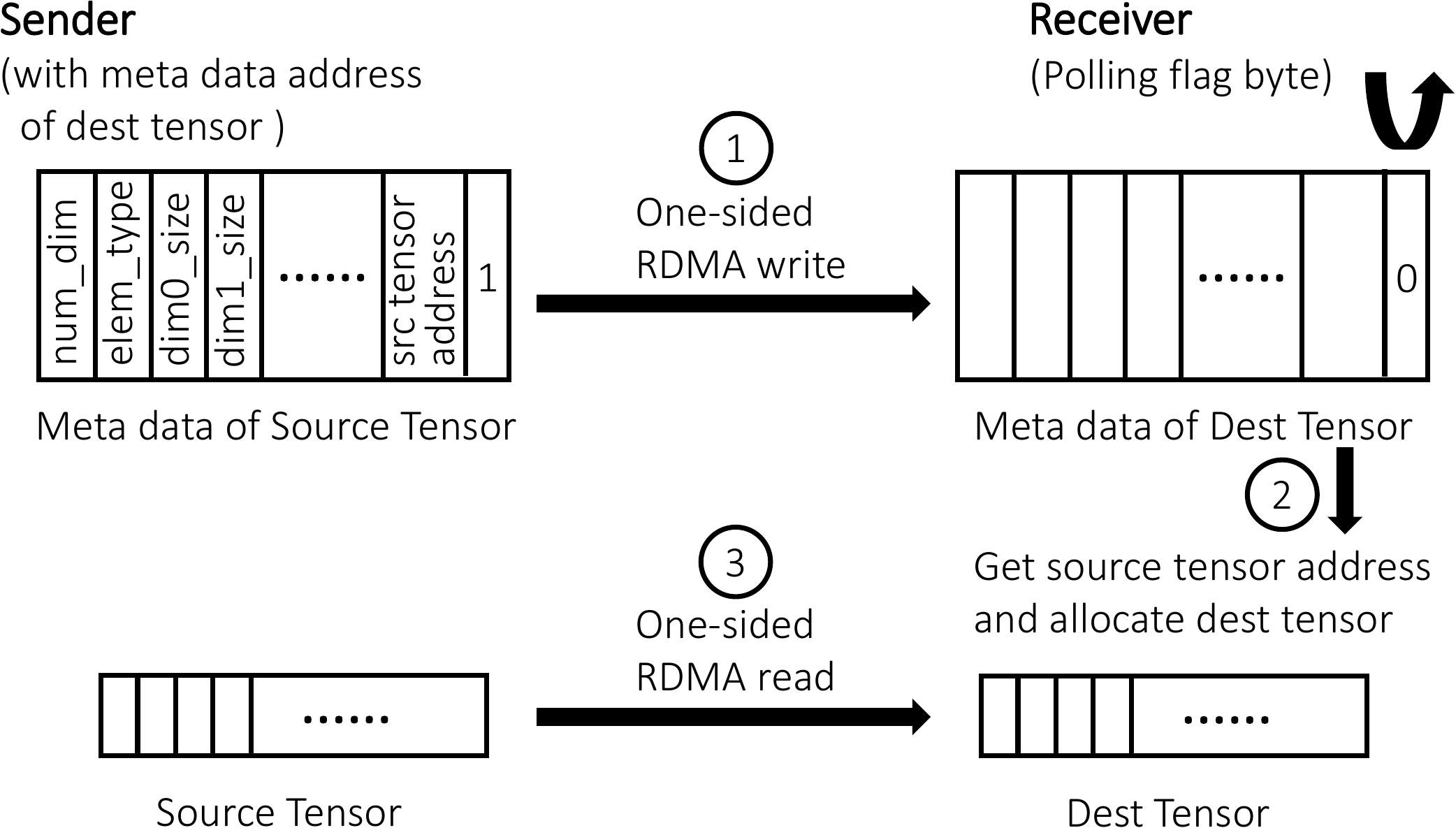}
	\vskip -2ex
	\caption{Transfer dynamically allocated tensor through one-sided RDMA write and read.}
	\vskip -2ex
	\label{fig:dymem}
\end{figure}

It is not always the case that the tensor placement can be decided statically. The shapes of the tensors that need to be transferred across servers can depend on the training data in each mini-batch iteration, and hence can change across different mini-batches. This is often the case where the deep learning applications have the training dataset with sparse features; e.g., the RNN model for natural language processing~\cite{seq2seq14} with input sequences having variant length in different mini-batches, and the wide-and-deep model used for recommender systems~\cite{widedeep16} with each training sample containing a different set of features. 


For these cases, although the graph analysis engine cannot fix the tensor placement during the computation, we still follow our design principle to reduce the communication-related computation overhead in a best effort. We observe that, despite the variance of tensor shape, the number of dimensions of a tensor is unchanged during the computation. Fixed tensor dimension count means that the size of the meta-data of a tensor is unchanged. With this assumption, we adapt the tensor transfer mechanism as illustrated in Figure~\ref{fig:dymem}. 

As shown in the figure, the meta-data includes the number of dimensions, the size of each dimension, the element data type of the tensor, and the remote address of the sender side tensor. The meta-data of the tensor at the receiver side is preallocated and its address is distributed to the sender side before the computation. During the computation, the sender writes the receiver side meta-data through the memory copy interface when the sender side tensor is ready to use. The receiver polls the flag byte at the tail of the meta-data. Once it detects the completion of the meta-data writing, it clears the flag byte, allocates a new tensor storage in the RDMA accessible memory region, and issues a remote memory copy to transfer the tensor data through one-sided RDMA read. Compared with the case of transferring statically placed tensor, the mechanism for passing the dynamically allocated tensor incurs the overheads of tensor allocation and the meta-data serialization and transfer. 


\input{analysis}

\input{gdr}

%% file: analysis.tex
\subsection{RDMA-Aware Graph Analysis}
\label{sec:analysis}
Given the mechanisms of the tensor transfer across servers through direct memory access, the data-flow graph analyzer can be enhanced to collect and provide useful information to make communication more efficient. First, for each tensor that need to be transferred through network, the analyzer needs to decide whether its shape can be known statically and never changes during the entire computation. This can be achieved through user annotations on the input tensors of the data-flow graph together with the shape-inference function of each graph node which tells the shapes of output tensors of the node given the shapes of its input tensors. With such information, a simple static analysis on the graph helps infer the shapes of some tensors. Second, after the graph is partitioned onto different servers, and on each server, before the sub-graph is executed, the data buffers of the receiver-side tensors whose shapes can be statically decided (or their meta-data buffers otherwise) are preallocated. The remotely accessible addresses of these tensor buffers are then passed to the servers holding the upstream tensors that they depend on. The delivered addresses are then set associated with the corresponding sender-side graph nodes which are responsible for transferring these tensors. 

On the sender side, a memory buffer that is to be transferred through RDMA needs to be registered beforehand to the RDMA NIC to allow its access. This registration process involves OS kernel actions such as pinning the buffer as non-pageable and introduces extra overhead. In addition, the allowed number of registered buffers is bounded by the specific RDMA hardware. Therefore, simply registering the data buffer of each tensor on demand when it needs to be transferred to remote server could introduce significant overhead and might experience unexpected error due to hardware resource limit. A more appropriate way of managing the RDMA-accessible memory is to preallocate a relatively large memory buffer to register to RDMA NIC once and employ a memory allocator on top of it for allocating smaller buffers for transfer. 

Normally, a sender of a tensor through RDMA would need to allocate an extra RDMA-accessible buffer and copy the tensor from the original buffer to it.
To avoid this memory copy, the graph analyzer would prefer allocating the RDMA-accessible buffer directly for the to-be-transferred tensor. 
One challenge to achieve this is to find out when a specific tensor is allocated (or its allocation site): the actual storage of an input tensor of a graph node might not be allocated at the execution of its direct predecessor node, because some graph nodes may conduct in-place manipulation on their input tensors, and hence a tensor buffer may be passed through multiple nodes on a path in the graph. We therefore propose a dynamic analysis method to address this. 

In order to get the allocation site of the storage of a tensor that is to be transferred, the graph analyzer instruments the tensor allocator used in the graph execution runtime. During the execution of the graph for the first mini-batch iteration, for each tensor allocation, it records the data buffer address of the tensor and the information of the corresponding graph node that invokes this allocation into a map with the tensor buffer address as the key. This node information includes the identification of the graph node and the Id of the allocation of this node, e.g., the \emph{i}th invocation of allocation in the execution of the node. If the information with the same address already exists in the map, the new information overwrites the old one. This way, we always keep the latest information with the same tensor address. 
When the graph node transfer a tensor during the execution, the runtime gets the tensor buffer address and looks up the map to get the information of the graph node that allocates the tensor buffer. 
It then stores the information of the tensor-allocating graph node into the set $S$ of memory regions that should ideally be allocated in the RDMA-accessible region directly. 
During the graph execution of the subsequent mini-batches, for each tensor allocation, the runtime checks whether the executing graph node exists in set $S$. If so, it allocates a tensor buffer from the allocator that manages the RDMA-accessible memory regions; otherwise, it allocates the tensor buffer from the normal allocator. This way, the data buffer of a to-be-transferred tensor, captured in $S$, is naturally RDMA-accessible and without need of extra copy. 


%% file: gdr.tex
\subsection{GPUDirect RDMA}
\label{sec:gdr}
GPUDirect RDMA is a technology that allows RDMA NIC to directly access GPU memory, and hence achieve direct access GPU memory remotely without going through host memory,
further reduces memory copy when the to-be-transferred tensor data is in GPU memory.
With the design principle and methodology in our work, applying GPUDirect RDMA is straightforward because, at user-level, it similarly just needs to allocate a GPU memory space in a mapped pinned mode through CUDA API~\cite{cuda} and register to RDMA NIC, and the graph analyzer can decide which tensors need to be allocated in this way as described in \S\ref{sec:analysis}.

It is relatively tricky to efficiently poll a value in GPU memory. Issuing a GPU kernel for every time of polling at an address may incur much overhead of kernel launch, and using a kernel function to repeatedly poll the address till the state becomes ready will waste the precious GPU computing resources. We therefore always employ the mechanism with dynamic allocation described in Section~\ref{sec:dyn} for tensor transfer through GPUDirect RDMA. Specifically, the meta-data of a tensor can be maintained in host memory so the polling only happens at CPU side, while the actual tensor data can be stored in GPU memory and transferred through one-sided RDMA read.


%% file: imp.tex
\section{Implementation}
\label{sec:imp}

We implement our techniques in \tf{} (r1.2)~\cite{tensorflowsc}, a popular open-sourced deep learning framework in community and industry. Our implementation contains about 4,000 lines of C++ code, where the RDMA communication library (using the \texttt{libibverbs} API on Linux) takes about 1,800 lines and the rest are modifications to \tf{} including the graph analyzer.

\tf{} organizes a deep learning computation as a data-flow graph. Users first build a graph through its high-level Python or C++ interfaces and then initiate the deep learning computation through associating the graph with a runtime session. The graph is composed of tensors and operators.
The operators refer to the computing operations of the corresponding graph nodes, while the tensors represent data flowing through the edges connecting the nodes.
Users are also allowed to develop customized operators and add those into the graph. Unlike the normal computational operators that are added in graph by users during the graph build phase, \texttt{Send} and \texttt{Recv} operators, which are used to transfer tensor data along edges across graph partitions, are added in the graph by the framework and are transparent to users. 


To implement the mechanisms of transferring tensor data over RDMA as described in \S\ref{sec:design}, we develop two pairs of custom operators and introduce an extended scheduling mechanism. For transferring tensors with a static placement, we implement the \texttt{RdmaSend} and \texttt{RdmaRecv} operators. During the graph analysis phase, the receiving tensor is preallocated with RDMA-accessibility and set as a property of \texttt{RdmaRecv}. This tensor is never freed until the entire computation finishes, so its address never changes in the entire computation. The remote-accessible address of the tensor is then passed to the server that holds the corresponding \texttt{RdmaSend} operator and set as its property. Once \texttt{RdmaSend} is scheduled to execute, it directly updates the content of the receiving tensor through a one-sided RDMA write. There is no need for some special mechanism to notify \texttt{RdmaSend} that the transferred tensor has been consumed by \texttt{RdmaRecv} because the next scheduled execution of \texttt{RdmaSend} is naturally guaranteed to happen after the consumption of the received tensor due to the control dependency of the loop in the graph or the execution sequentially of multiple mini-batch iterations. Similarly, we also implement another pair of operators, \texttt{RdmaSendDyn} and \texttt{RdmaRecvDyn} for supporting the tensor transfer with dynamic allocation as described in \S\ref{sec:dyn}.

\tf{} originally have 
two types of execution modes for operators: synchronous and asynchronous execution. For both types of operators, once an operator is popped out of the ready queue to execute, it simply completes its execution synchronously or asynchronously without the need to be enqueued into the ready queue again. However, the \texttt{RdmaRecv} and \texttt{RdmaRecvDyn} need to poll the flag byte in the data or meta-data buffer of the receiving tensor. If executing totally away from the scheduling mechanism, it either suffers from busy loop wasting processor resources or long latency due to periodic sleep. We therefore introduce a new execution mode of operator called \emph{polling-async}. The execution of this type of operator contains two phases. When executing in the polling phase, the scheduler checks whether the polling succeeds. If not, it simply re-enqueues this operator into the tail of the ready queue; otherwise, it changes the execution mode of the operator to asynchronous and reschedules the execution. This way, we reduce the polling  overhead when there are other ready operators to execute. 

%% file: eval.tex
\section{Evaluation}
\label{sec:eval}


\begin{table}[t]
	\centering
	\footnotesize
	\begin{tabular}{l|l|r|c|r}
		Type & Benchmark & \makecell{Model size \\ (MB)} & \makecell{Variable \\ Tensor\#} & \makecell{Computation \\ time (ms)}\\ \hline
		\multirow{3}{*}{CNN} & AlexNet 		& 176.42 	& 16 	& 7.61 $\pm$ 0.29 \\ 
		& Inception-v3 & 92.90 	& 196	& 68.32 $\pm$ 0.73 \\
		& VGGNet-16 		& 512.32 & 32 	& 30.92 $\pm$ 0.19 \\ 
		\hline
		\multirow{2}{*}{RNN} & LSTM 		& 35.93 	& 14	& 33.33 $\pm$ 0.24 \\ 
		& GRU 			& 27.92 	& 11	& 30.44	$\pm$ 0.32 \\ \hline
		\multirow{1}{*}{FCN} & FCN-5 		& 204.47 & 10 	& 4.88 $\pm$ 0.28 \\ 							 
	\end{tabular} 
	\vskip -1ex
	\caption{Deep learning benchmarks (Note: the LSTM and GRU are configured with hidden vector size of 1024 and step size of 80; the FCN-5 consists of 3 hidden layers with dimension of 4096 and two layers of input and output)}
	\vskip -2ex
	\label{tab:bench}
\end{table}

\begin{figure}
	\centering
	\includegraphics[width=0.75\linewidth]{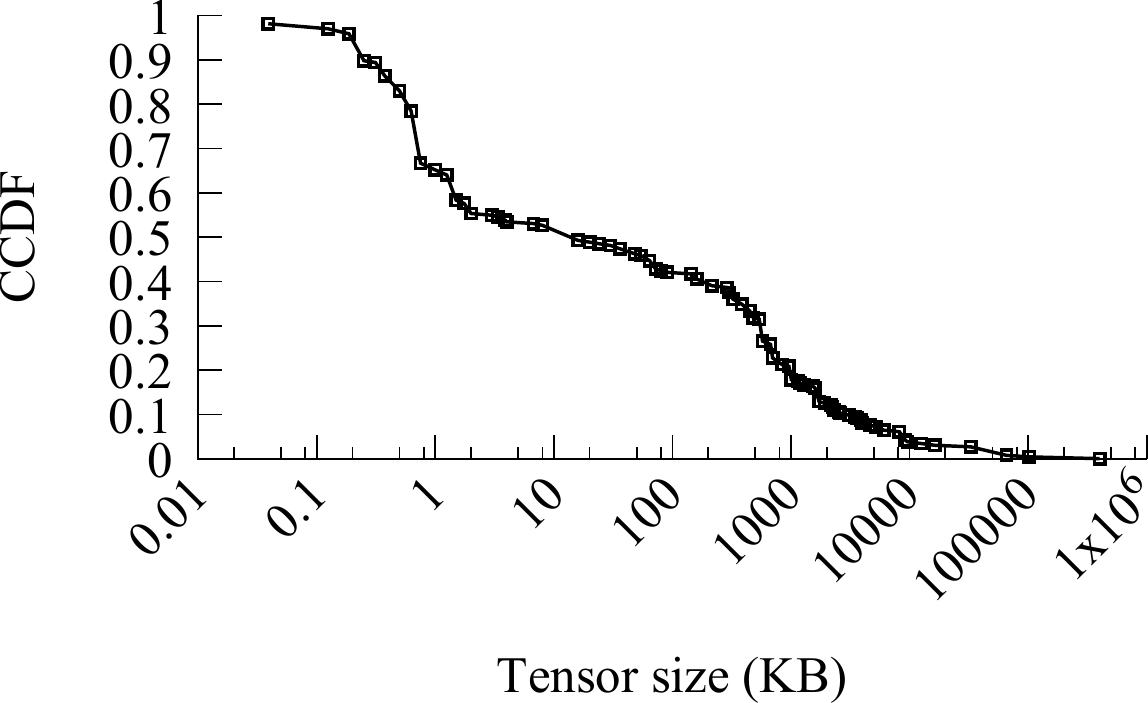}
	\vskip -2ex
	\caption{The complementary cumulative distribution of variable tensor sizes}
	\vskip -3ex
	\label{fig:ccdf_tensor}
\end{figure}

We evaluate our techniques on a cluster that consists of 8 servers. Each server is equipped with dual $2.6\,${GHz} Intel Xeon E5-2690v4 14-core CPU, $512\,${GB} memory, 2 NVIDIA Tesla P100 GPU, and a $100\,${Gbps} Infiniband (IB) network adapter (Mellanox MT27700) for interconnection. All the servers are installed with Ubuntu 16.04, CUDA 8.0, and cuDNN 6.
\tf{} (since version r1.0) supports RDMA in the way of wrapping RDMA communication layer with the gRPC abstraction, and hence has to maintain private message buffers and introduce extra memory copy.
It also relies on some IB-specific features and can only run on an IB cluster, while our RDMA mechanism can also work with RoCE (RDMA over Converged Ethernet) network adapters.

We extensively evaluate the performance with a set of representative deep learning benchmarks, including Inception-v3~\cite{inception15}, AlexNet~\cite{alexnet12nips}, VGGNet-16~\cite{vggnet14}, LSTM~\cite{lstm97}, GRU~\cite{grucho14} and FCN-5, covering convolutional neural network (CNN), recurrent neural network (RNN) and fully connected neural network (FCN). Table~\ref{tab:bench} lists some characteristics of these benchmark workloads. The model size is the sum of the sizes of all the variable tensors in the neural network, which corresponds to the communication volume between workers and parameter server processes in each mini-batch. The local computation time represents the average execution time of processing one sample data in the single-server setting. We therefore use the model size of each benchmark to characterize its network load and use the local computation time to characterize its computation complexity. These benchmarks cover both computation-intensive and network-intensive workloads. For example, the Inception-v3 model is a typical computation intensive workload, while the VGGNet-16 is mainly bottlenecked in network, because each worker needs to transfer more than 1~GB ($2\times512.32$MB) model and gradient data in each mini-batch.
Among these benchmarks, the sizes of variable tensors vary from tens of bytes to hundreds of megabytes, however, in many cases, the existence of large tensors may substantially influence communication behavior. Figure~\ref{fig:ccdf_tensor} shows the distribution of number of tensors with different tensor sizes in our benchmark. As shown in the figure, more than $50\%$ of the variable tensors are larger than 10KB, and more than $20\%$ are even larger than 1MB. In terms of the total capacity, the tensors that are larger than 1MB occupy $96\%$ of the capacity among all tensors.


We conduct most of the experiments on synthetic datasets that are randomly generated and mainly used for evaluating the execution time. To demonstrate the effect of our techniques in real scenarios, we also evaluate the convergence of 3 end-to-end applications on a real-world datasets, which includes a translation task  (Seq2Seq) using the sequence-to-sequence model~\cite{seq2seq14} on WTM'10 French-English machine translation corpus~\cite{wtm15} containing about 20~GB text data in total, an image recognition task (CIFAR) using the CIFAR-10 model on its public dataset~\cite{cifar_data} consisting of 60000 32$\times$32 colour images in 10 classes, 
and an RNN based sentence embedding task (SE) used in our real production. We use a private production dataset containing about 3.7~GB text data in this model.

All performance numbers with respect to throughput (mini-batches$/$second) in our experiments are calculated by averaging among 5 runs with each processing 100 mini-batch iterations. In all cases we observed very little variation, thus we omit the error bars in all figures.


\input{microbench}
\input{e2ebench}

%% file: microbench.tex
\subsection{Performance on Micro-benchmark}
\label{sec:micro}

\begin{figure}
	\centering
	\includegraphics[width=0.85\linewidth]{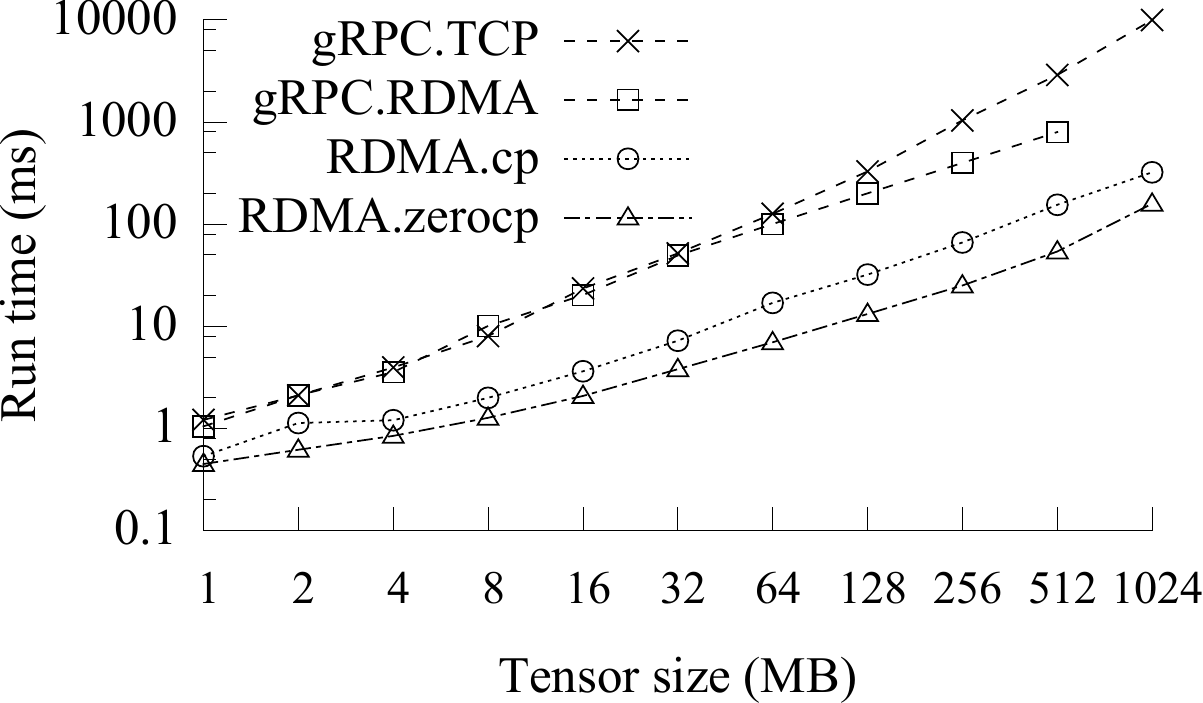}
	\vskip -2ex
	\caption{The performance comparison on send/receive micro-benchmark with two servers.}
	\vskip -2ex
	\label{fig:sendrecv_bench}
\end{figure}

In order to understand the direct benefit of our design on system performance, we first evaluate our tensor transfer mechanism over RDMA using a micro-benchmark. 

We set up two servers only to perform a tensor transfer, so as to compare our network performance with gRPC over TCP and gRPC over RDMA. The receiver also performs a lightweight \texttt{reduce\_max} operator to consume the passed tensor.
Figure~\ref{fig:sendrecv_bench} shows the efficiency of transferring tensors with different sizes. 
We first compare our RDMA-based mechanism (i.e., RDMA.zerocp) with the \tf{}'s original gRPC-based solutions, including both the gRPC over TCP (i.e., gRPC.TCP) and the gRPC over RDMA (i.e., gRPC.RDMA). As shown in the figure, our mechanism can outperform both of them significantly. For example, RDMA.zerocp can improve the speed by 1.7$\times$ to 61$\times$ than gRPC.TCP for different message sizes. 
For gRPC.RDMA, even though it adopts RDMA protocol under gRPC, it still needs to conduct data serialization/de-serialization and data copy between RDMA pinned buffer and tensor memory on both the sender and receiver sides. In contrast, our RDMA-based mechanism can completely avoid any data copy and serialization overhead, and hence gets 1.3$\times$ to 14$\times$ performance improvement compared to gRPC.RDMA for different message sizes. Note that, there is a missing data point for gRPC.RDMA at message size of 1GB, because \tf{} with gRPC.RDMA will crash when the transferring data size is larger than 1GB.
To evaluate the memory copy overhead, we manually turn off our graph analysis optimization, so that the tensor data in the sender side is unable to be pre-allocated as RDMA-accessible. To perform tensor transfer, the \texttt{RdmaSend} operator has to allocate a new RDMA-accessible buffer, copy the tensor into it, and then conduct the actual RDMA write. The curve of RDMA.cp in Figure~\ref{fig:sendrecv_bench} demonstrates the performance of this case. As it shows, RDMA.zerocp outperforms the RDMA.cp by 1.2$\times$ to 1.8$\times$ for different message sizes. Note that this improvement is far less than the gap between gRPC.RDMA and RDMA.zerocp, because RDMA.cp mechanism only involves the data copy on the sender side and does not involve any data serialization/de-serialization overhead.

%% file: e2ebench.tex
\subsection{Performance on Deep Learnings}
\label{sec:perfdnn}

This section evaluates our system on real deep learning applications.
Benchmarks listed in Table~\ref{tab:bench} are evaluated on synthetic data for performance, while the 3 aforementioned applications on real datasets are evaluated for convergence.
By default, experiments are configured as running in distributed settings with data-parallelism, where each machine runs a worker process and a parameter server process. During execution, each worker executes a data-flow graph replica on a portion of training data. The variable tensors are shared across workers and are placed in parameter servers in a round-robin fashion. The worker runs multiple iterations
until some convergence condition is satisfied or a maximum iteration number is reached. In the following discussion, unless stated explicitly, we always compare our fully-optimized RDMA mechanism with other alternative solutions.

\paragraph{Performance.}


\begin{figure*}[h]
	\centering
	\subfigure[Alexnet]{
		\includegraphics[width=0.65\columnwidth]{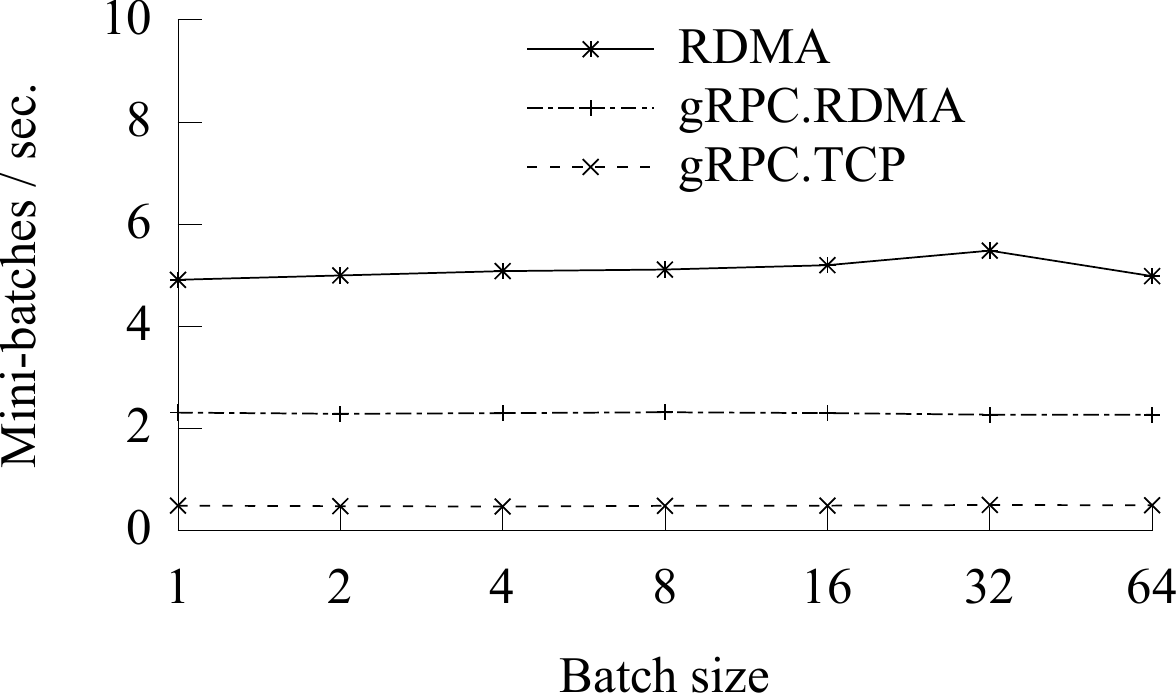}
	}
	\subfigure[Inception-v3]{
		\includegraphics[width=0.65\columnwidth]{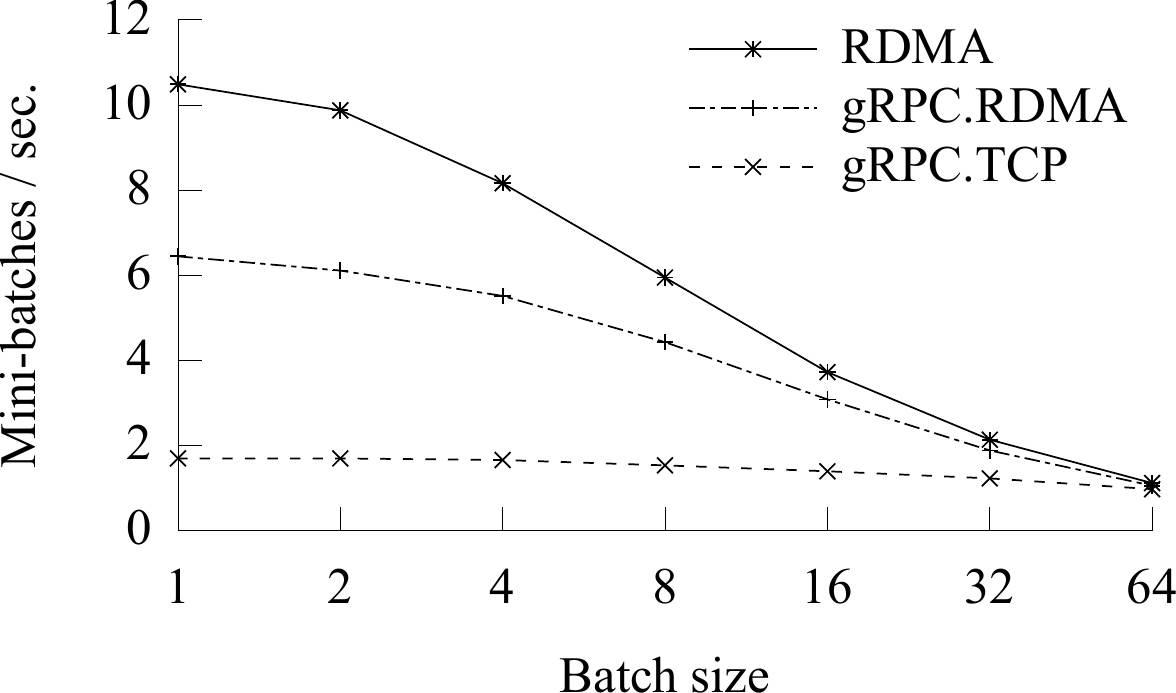}
	}	
	\subfigure[VGGNet-16]{
		\includegraphics[width=0.65\columnwidth]{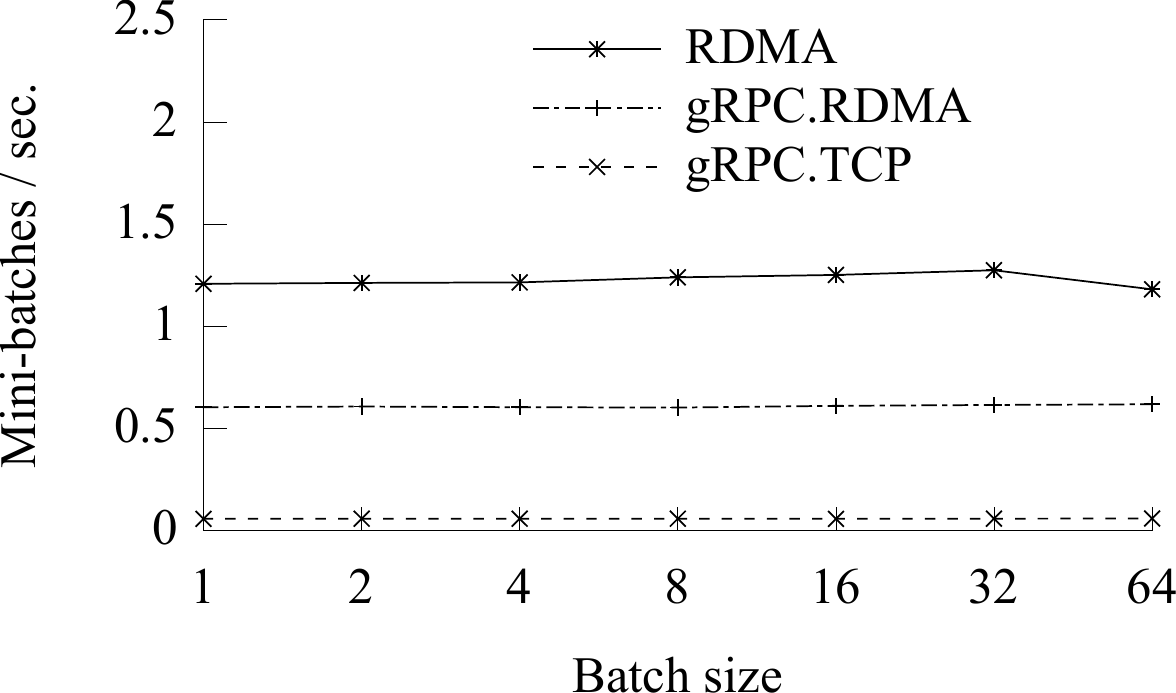}
	}
	\subfigure[LSTM]{
		\includegraphics[width=0.65\columnwidth]{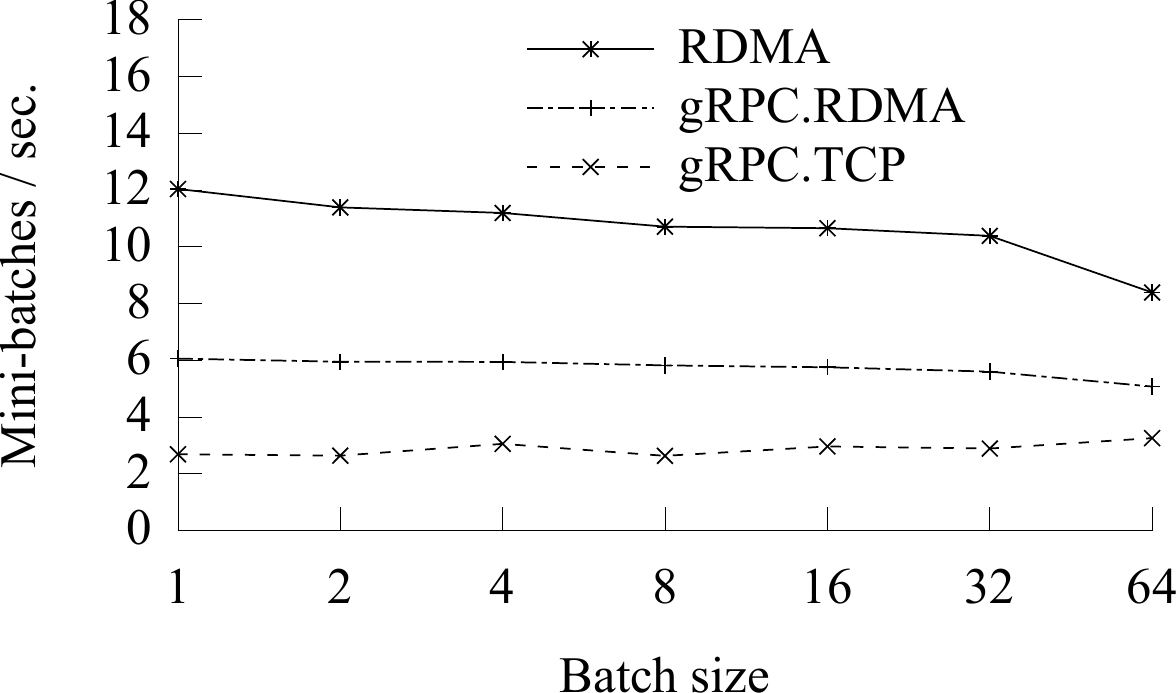}
	}
	\subfigure[GRU]{
		\includegraphics[width=0.65\columnwidth]{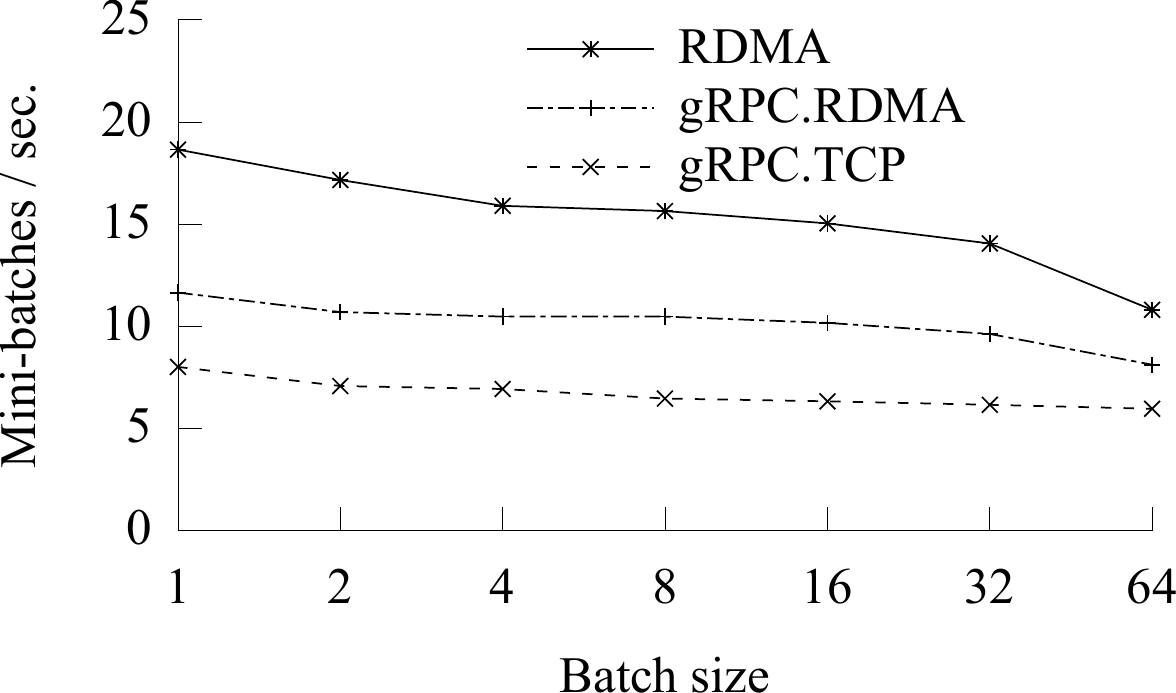}
	}	
	\subfigure[FCN-5]{
		\includegraphics[width=0.65\columnwidth]{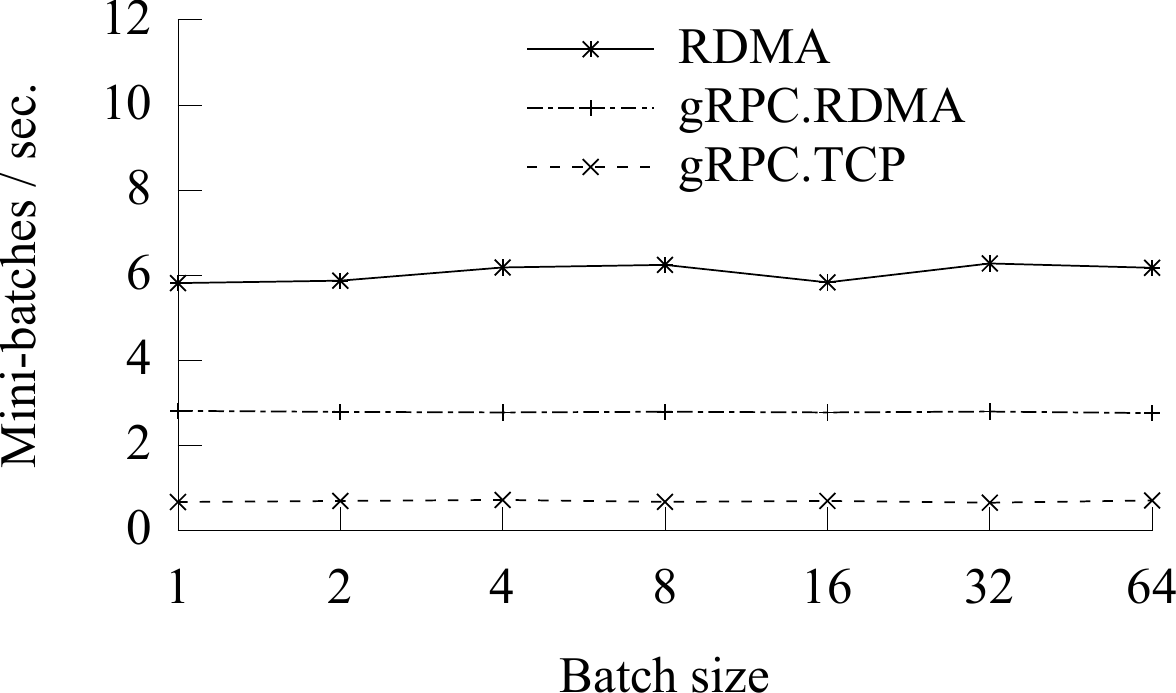}
	}
	\vskip -2ex
	\caption{Comparison with gRPC-based solutions in TensorFlow.}
	\vskip -3ex
	\label{fig:e2e-comp}
\end{figure*}

We run the 6 deep learning benchmarks with a synthetic dataset on the same cluster. To understand better the computation and communication behavior, our synthetic datasets are generated on the fly, which can avoid the overhead of data loading from disk. In deep learning applications, the mini-batch size is a critical hyper parameter that affects both the convergence rate and the computation time. In distributed training, we can amortize the communication overhead by using large batch size, because it can increase the local computation time. However, a large batch size is often harmful for convergence, because it reduces the model synchronization frequency across different workers. In practice, the optimal setting is tuned by users, but the common batch sizes are around the numbers like 16, 32, or 64. In our experiments, we evaluate each benchmark with batch sizes ranging from 1 to 64 (128 for some).

Figure~\ref{fig:e2e-comp} plots the performance of \tf{} with gRPC and with our RDMA mechanism. 
In general, the average improvements from using RDMA against gRPC.RDMA 
range from 117\% up to 145\% for VGGNet-16. 
The improvements observed for other benchmarks reach: 
169\% for AlexNet, 65\% for Inception-v3, 151\% for FCN-5, 118\% for LSTM, and 69\% for GRU.
And the improvements over gRPC.TCP is much greater; for example, 25$\times$ for VGGNet-16.

As shown in the figure, among these benchmarks, AlexNet, VGGNet-16, and FCN-5 get relatively more significant improvements from RDMA than others, because they are mainly bottlenecked at communication. Their execution time is stable under different batch sizes, because the volume of transferred data (i.e., the model size) is irrelevant to the batch size and the GPU's massive computing threads can complete large batches within the same time as processing the small ones. However, for other benchmarks like the Inception-v3, LSTM, and GRU, when we increase the batch size to larger than 32, their local computation time also increases and becomes dominant in the overall execution time. In those cases, the gaps between gRPC and our RDMA decrease as expected.

\paragraph{Convergence.}


\begin{figure*}[h]
	\centering
	\subfigure[Seq2Seq]{
		\includegraphics[width=0.65\columnwidth]{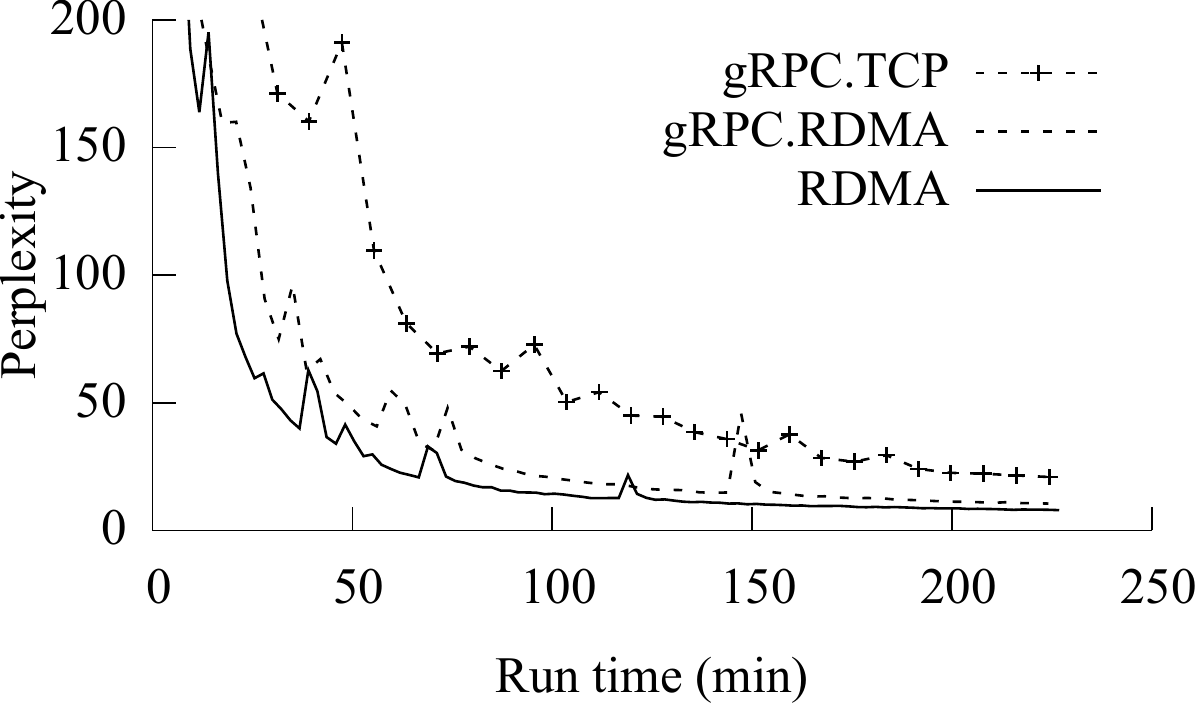}
	}
	\subfigure[CIFAR]{
		\includegraphics[width=0.65\columnwidth]{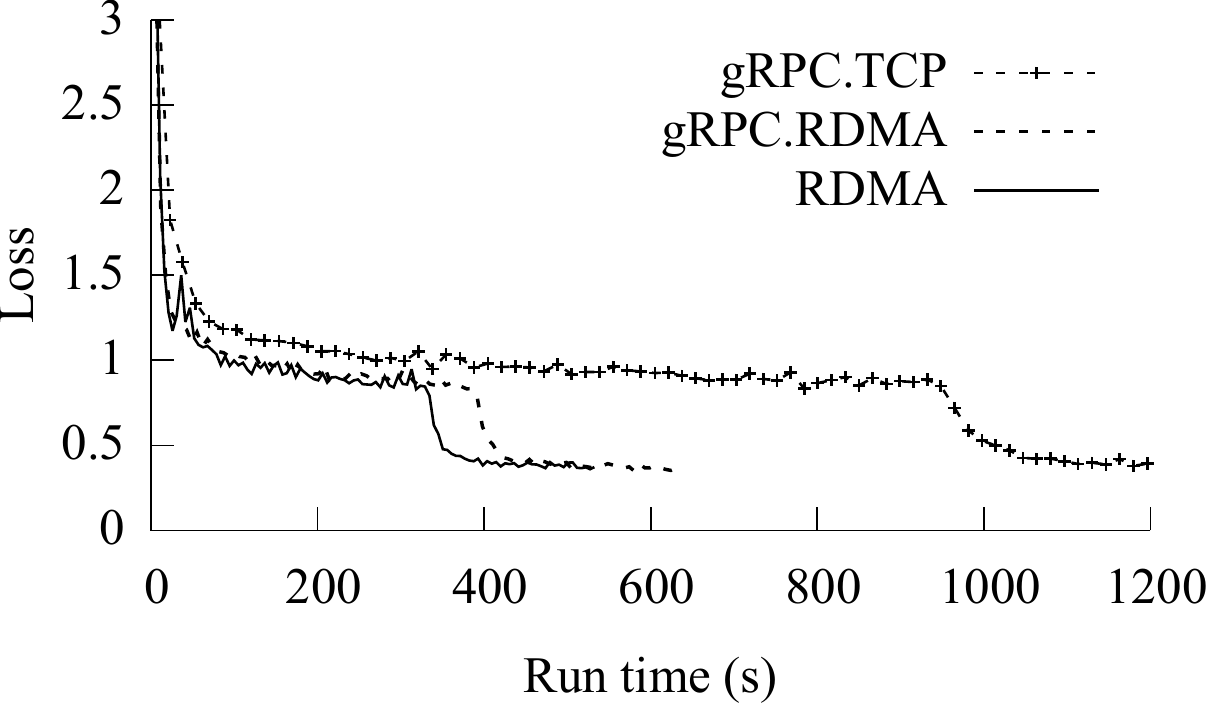}
	}	
	\subfigure[SE]{
		\includegraphics[width=0.65\columnwidth]{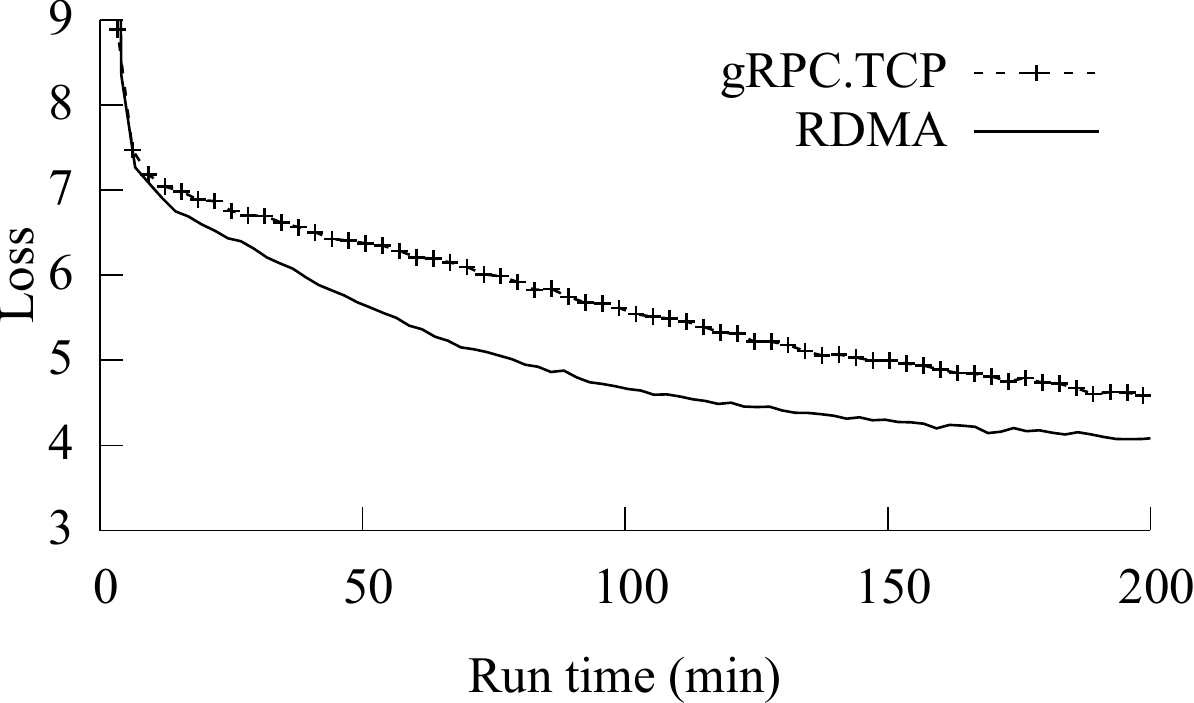}
	}
	\vskip -2ex 
	\caption{Convergence of real applications on \tf{} with different communication mechanisms.}
	\vskip -2ex 
	\label{fig:converge}
\end{figure*}

To demonstrate the performance gain in real scenarios, we further evaluate   three end-to-end training tasks, including a translation task (Seq2Seq) using the  sequence-to-sequence model~\cite{seq2seq14}, an image recognition task (CIFAR) using the CIFAR-10 model~\cite{cifar_data} and a sentence embedding task (SE) based on two RNN models. We use perplexity value for Seq2Seq model and loss value for others to measure the convergence quality.
For each task, we randomly partition their dataset into 8 workers. Each worker continuously loads the sample data from local disk in parallel with the training process.
For each model, we compare gRPC.TCP, gRPC.RDMA, and our RDMA mechanism on the same training dataset until convergence.

Figure~\ref{fig:converge} plots the convergence curves for the three models with different communication mechanisms. For the Seq2Seq model in Figure~\ref{fig:converge}(a), it takes about 220 minutes to converge to perplexity under 20 with gRPC.TCP. However, when using our RDMA mechanism, it takes only 66 minutes, about 3$\times$ speedup. Even comparing to gRPC.RDMA, our RDMA mechanism achieves 53\% performance improvement.
Similar results can be observed in the CIFAR model (Figure~\ref{fig:converge}(b)) and the SE model (Figure~\ref{fig:converge}(c)). For the CIFAR model, our RDMA mechanism can speed up convergence by 2.6$\times$ compared to gRPC.TCP, and 18\% to gRPC.RDMA. 
Finally, for the SE model, we fail to collect the results of gRPC.RDMA because \tf{} crashes when using gRPC.RDMA.
If just using gRPC.TCP, the SE model can converge to loss value of 4.5 within 185 minutes. However, our RDMA mechanism takes only about 100 minutes to converge to the same point, which speeds up the training process by 85\% in total.

\begin{figure*}[h]
	\centering
	\subfigure[LSTM]{
		\includegraphics[width=0.65\columnwidth]{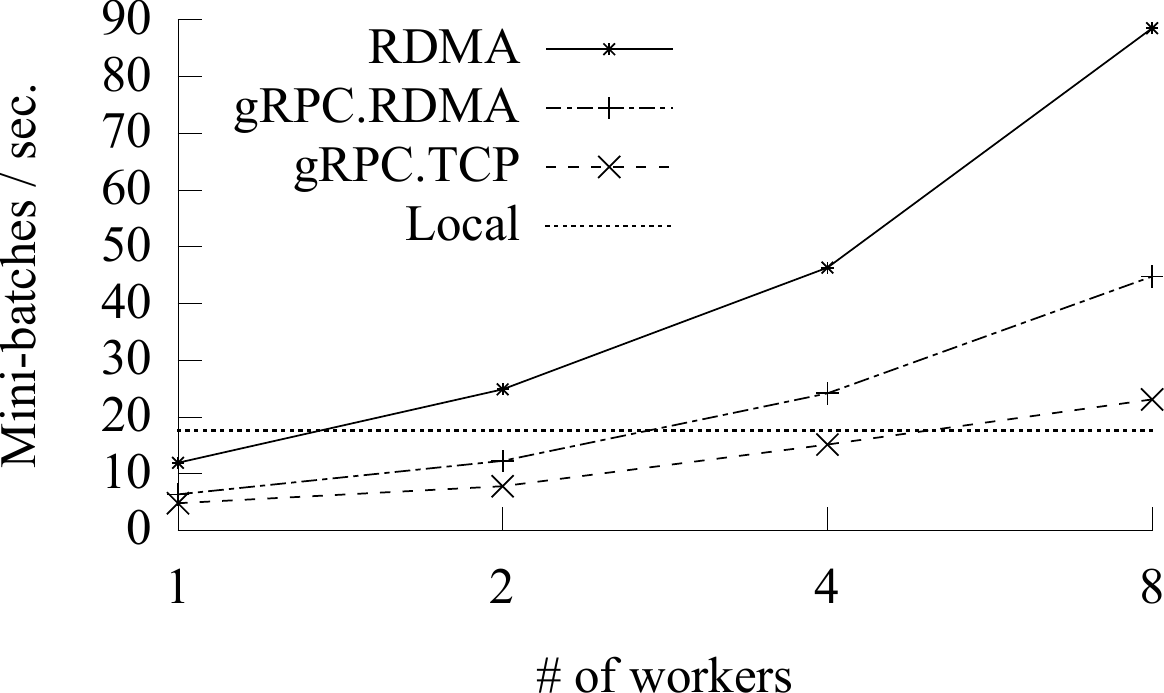}
	}
	\subfigure[Inception-v3]{
		\includegraphics[width=0.65\columnwidth]{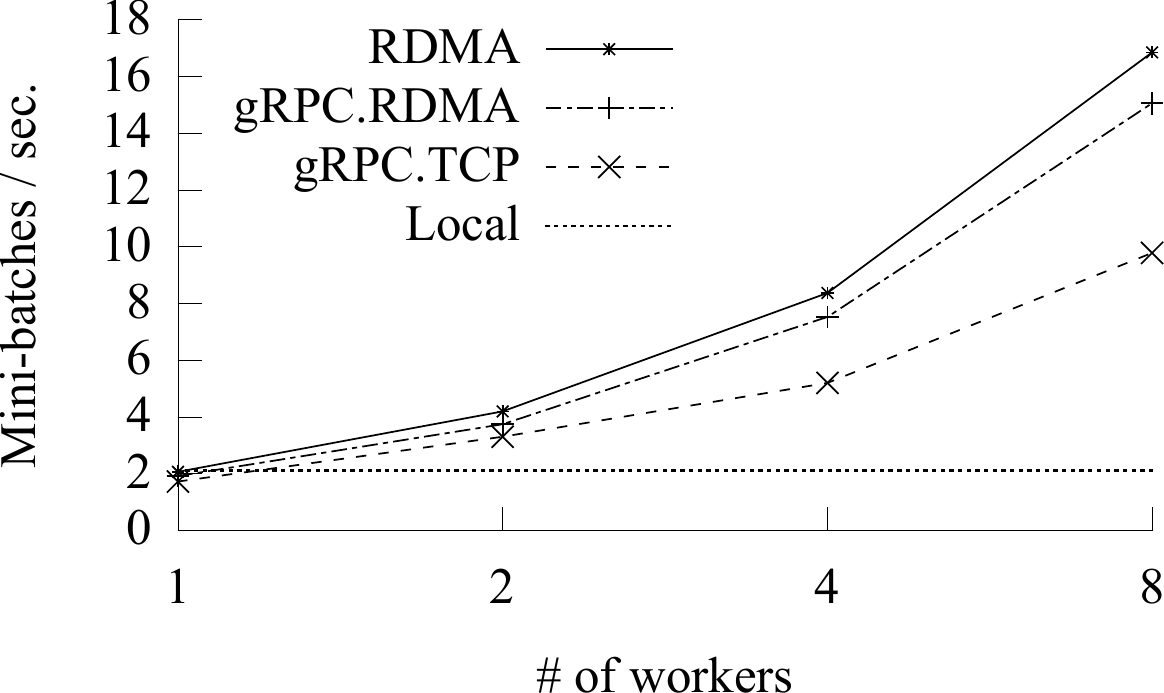}
	}	
	\subfigure[VGGNet-16]{
		\includegraphics[width=0.65\columnwidth]{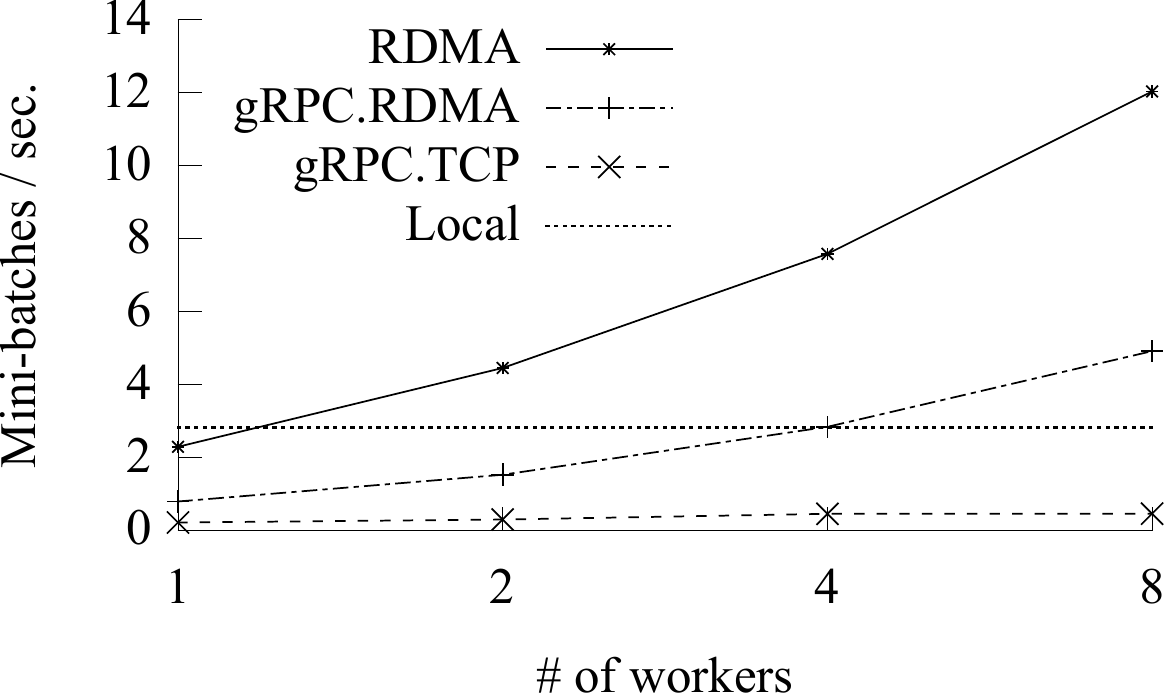}
	}
	\vskip -2ex
	\caption{Scalability of \tf{} with gRPC-based solutions vs. RDMA.}
	\vskip -2ex
	\label{fig:scalability}
\end{figure*}


\paragraph{Scalability.}
Scalability is one of the most important metrics for distributed training. We evaluate the scalability of \tf{} with both our RDMA mechanism and the original solutions on all the deep learning benchmarks with synthetic datasets. 
We set batch size as 32 for all experiments.

The scalability of each benchmark is mainly determined by its computation and communication behavior. Figure~\ref{fig:scalability} shows the scalability results of three representative workloads among all the benchmarks. We can see that different benchmarks have very different scalability patterns. For the LSTM and Inception-v3, because they are computation intensive at batch size 32, we can always observe good scalability no matter whether we use our RDMA mechanism or gRPC. For example, the speedup on 8 servers for both RDMA solutions on the two benchmarks are larger than $7\times$ against their single server cases (still involving communication between workers and parameter server processes on the same machine). Even in those cases, our RDMA remains much better than gRPC based RDMA in terms of throughput (i.e., 98\% higher for LSTM and 12\% for Inception-v3). 
For VGGNet-16, because it is a communication intensive application, its scalability is highly determined by the underlying network efficacy. In this case, our RDMA can still get 5.2$\times$ speedup against its single server, always holds more than 140\% faster than gRPC based RDMA on different scales.

For each of these benchmarks, we also measure the throughput of its pure local implementation (the ``Local'' line in Figure~\ref{fig:scalability}), which does not involve any communication overhead.
As shown in the figure, for the gRPC.RDMA case, the speedups on 8 servers relative to the local implementations for LSTM and Inception-v3 are 1.5$\times$ and 6$\times$, respectively.
It needs 4 servers to outperform the local implementation for LSTM and 
even with 8 servers for VGGNet-16 due to its much more serious communication bottleneck.
In contrast, with our RDMA, all the three distributed benchmarks can outperform the local implementations with only 2 servers. The speedups on 8 servers are 5$\times$, 7.9$\times$ and 4.3$\times$ for the three benchmarks, respectively.

\paragraph{Memory Copy Overhead.}

\begin{figure}[h]
	\centering
	\includegraphics[width=0.8\linewidth]{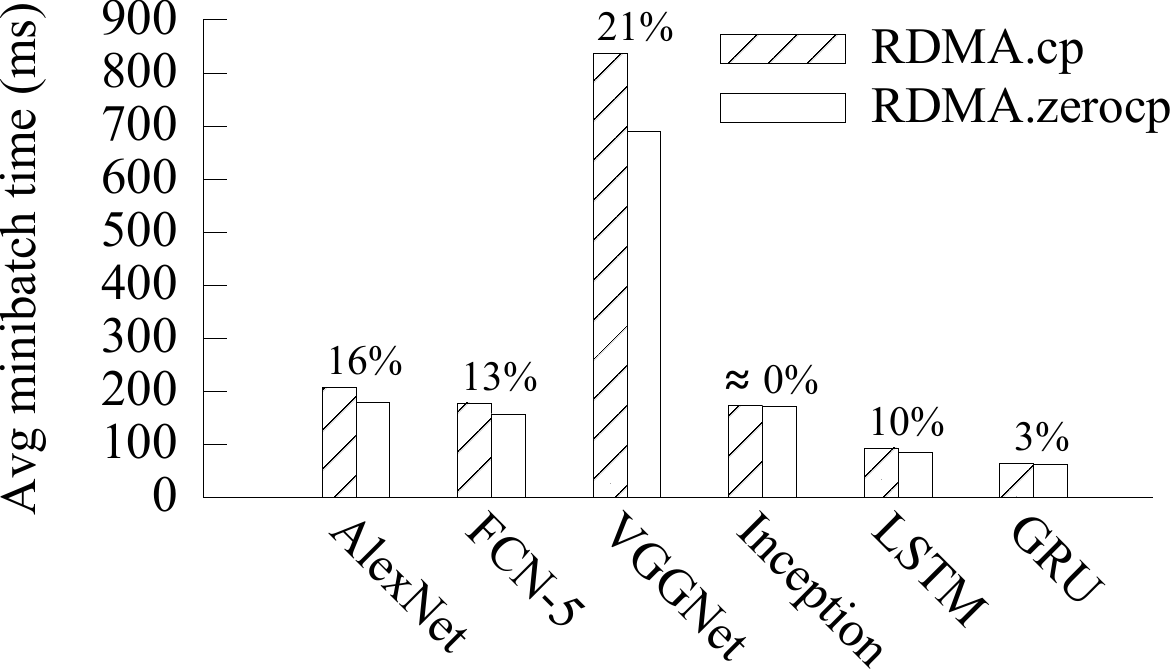}
	\vskip -2.5ex
	\caption{The memory copy overhead in deep learning benchmarks.}
	\vskip -1.5ex
	\label{fig:memcpy}
\end{figure}

We also evaluate the performance gain of removing the sender side memory copy, as described in Section~\ref{sec:micro}, in deep learning benchmarks. We manually turn off the graph analysis phase so as to skip the optimization for zero copy, and compare its performance with the optimized one. Figure~\ref{fig:memcpy} shows the average mini-batch time of each benchmark with (or without) memory copy. In general, for different workloads, the zero-copy optimization can bring up to 21\% performance improvement with mini-batch size of 8. However, for some benchmarks like the Inception-v3 and GRU, the performance gain is relatively small. This is mainly due to two factors. First, as we explained before, these benchmarks are mostly computation intensive, which could benefit little from network related optimization. Second, from the result of micro-benchmark in Section~\ref{sec:micro}, the gain of zero copy is more significant for larger tensor size, however, the Inception-v3 includes many small tensors since it has 196 variables but the total model size is only 92.9MB.

\paragraph{GPUDirect Support.}


\begin{table}[]
	\centering
	\footnotesize
	\begin{tabular}{l|l|l|l}
		Benchmark & RDMA  & RDMA+GDR & Improv. \\ \hline
		AlexNet   & 178.5 & 135.2    & 32\%    \\ \hline
		FCN-5     & 157.0 & 101.9    & 54\%    \\ \hline
		VGGNet    & 690.1 & 610.4    & 13\%    \\ \hline
		Inception & 172.5 & 171.9    & 0.4\%     \\ \hline
		LSTM      & 84.4  & 68.1     & 24\%     \\ \hline
		GRU       & 62.3  & 52.6     & 19\%    
	\end{tabular}
	\vskip -2ex
	\caption{The average minibatch time (ms.) and improvements with GPUDirect RDMA(GDR) in deep learning benchmarks. (8 workers)}
	\vskip -2ex
	\label{tab:gdr}
\end{table}

Finally, we evaluate the performance with GPUDirect RDMA enabled for different applications, as shown in Table~\ref{tab:gdr}. After enabling GPUDirect, our RDMA further improves the performance by up to 54\%. Improvements vary from application to application, similar to what we observed in previous experiments. 

%% file: related.tex
\section{Related Work}
\label{sec:related}
\paragraph{Systems leveraging RDMA.}
With the advent of the emerging RDMA technology, a large body of research has been done on improving the performance of various distributed systems to leverage its low latency and high bandwidth. A series of efforts~\cite{Jinyang13,farm14,herd14,rdmahtm15,nessie15} target to optimize key-value storage systems, while some others focus on improving the throughput of distributed transaction processing systems~\cite{rdmahtm15,rdmahtm16,compro15}. FaRM~\cite{farm14} uses one-sided RDMA reads for key-value lookups while employs a messaging primitive for updates. This messaging mechanism uses a fixed ring buffer on the receiver side to hold received messages, and hence may bring extra overhead of copying messages to the application buffers. In addition, large messages may have to be split on the sender side and re-assembled on the receiver side due to the limited size of the ring buffer, which introduces more copying overhead. HERD~\cite{herd14} embraces an RPC abstraction for key-value lookup to avoid multiple remote accesses on a hash-table structure. Kalia \emph{et al}.~\cite{rdma16,fasst16} further improves it by considering lower-level factors in RDMA hardware and optimizing it for the all-to-all cases used in transaction processing. All these research efforts target the scenarios dominated by small messages, where latency is the major objective of optimization.

Grappa~\cite{Grappa15} and GraM~\cite{Gram15} explore the use of RDMA to accelerate distributed graph computation. In graph processing, it is appropriate to use RPC to batch many small random remote accesses caused by the complex and sparse graph structure. However, in the deep learning scenarios, a common pattern is to access dense tensors, which are often relatively large.

GPUNet~\cite{gpunet14} proposes a socket-like abstraction over RDMA, which allows GPU kernel functions to directly communicate through network. Their design targets scenarios of general distributed computations on GPUs. It is interesting to look at how to integrate this level of ``direct'' into a dataflow-based deep learning framework combined with techniques in our work.

\paragraph{Distributed machine learning systems.} 
Many systems have been designed to support efficient distributed computation of traditional machine learning algorithms, which usually employ shallow model structure and do not necessarily express their computation as data-flow graph, such as Petuum~\cite{petuum15} and Li \emph{et al}.'s ``Parameter Server''~\cite{ParameterServerLi2014}. These shallow model structures often lead to sparse matrix computations, which share the similar patterns as graph processing~\cite{tux2017}. These systems scale out by employing a parameter server architecture, which uses a set of servers to manage shared state that is updated by a set of parallel workers. This parameter server architecture can also be used to support some distributed deep learning frameworks, such as DistBelief~\cite{disbelief12}, Project Adam~\cite{projadam14}, MxNet~\cite{mxnet15}, and so on. Although these frameworks, unlike \tf{}, only use data-flow graph to represent local computation at each worker, the principle and methodology in our work can also be applied in them to further improve their communication efficiency and scalability.

%% file: conclusion.tex
\section{Conclusion}
\label{sec:con}
The emerging deep learning workloads and network technologies such as RDMA have prompted us to rethink the widely used RPC abstraction for network communication. 
We observe that the abstraction does not allow application-level information to be passed to the network layer for optimizations,
leading to unnecessary additional memory copy and significant performance penalty.
By designing a simple ``device''-like interface, along with a combination of static analysis and dynamic tracing , we have enabled cross-stack optimizations for general deep neural network training to take full advantage of the underlying RDMA capabilities, leading to up to almost an order of magnitude speedup for representative deep learning benchmarks over the default RPC library and up to 169\% improvement even over an RPC implementation optimized for RDMA.

%% file: main.bbl
\begin{thebibliography}{10}

\bibitem{cuda}
{CUDA Driver API}.
\newblock \url{http://docs.nvidia.com/cuda/cuda-driver-api}.

\bibitem{grpc}
{gRPC - An RPC library and framework}.
\newblock \url{https://github.com/grpc/grpc}.

\bibitem{tensorflowsc}
{TensorFlow}.
\newblock \url{https://github.com/tensorflow/tensorflow/tree/r1.2}.

\bibitem{thrift}
{The Apache Thrift}.
\newblock \url{http://thrift.apache.org}.

\bibitem{wtm15}
{WTM'15 Machine Translation Dataset}.
\newblock \url{http://www.statmt.org/wmt15}.

\bibitem{zmq}
{ZeroMQ}.
\newblock \url{http://zeromq.org/}.

\bibitem{tensorflow16}
{\sc Abadi, M., Barham, P., Chen, J., Chen, Z., Davis, A., Dean, J., Devin, M.,
  Ghemawat, S., Irving, G., Isard, M., Kudlur, M., Levenberg, J., Monga, R.,
  Moore, S., Murray, D.~G., Steiner, B., Tucker, P., Vasudevan, V., Warden, P.,
  Wicke, M., Yu, Y., and Zheng, X.}
\newblock {TensorFlow: A System for Large-Scale Machine Learning}.
\newblock In {\em 12th USENIX Symposium on Operating Systems Design and
  Implementation (OSDI 16)\/} (GA, 2016), USENIX Association, pp.~265--283.

\bibitem{rpc84}
{\sc Birrell, A.~D., and Nelson, B.~J.}
\newblock Implementing remote procedure calls.
\newblock {\em ACM Trans. Comput. Syst. 2}, 1 (Feb. 1984), 39--59.

\bibitem{mxnet15}
{\sc Chen, T., Li, M., Li, Y., Lin, M., Wang, N., Wang, M., Xiao, T., Xu, B.,
  Zhang, C., and Zhang, Z.}
\newblock Mxnet: {A} flexible and efficient machine learning library for
  heterogeneous distributed systems.
\newblock In {\em NIPS Workshop on Machine Learning Systems (LearningSys)\/}
  (2016).

\bibitem{rdmahtm16}
{\sc Chen, Y., Wei, X., Shi, J., Chen, R., and Chen, H.}
\newblock Fast and general distributed transactions using rdma and htm.
\newblock In {\em Proceedings of the Eleventh European Conference on Computer
  Systems\/} (New York, NY, USA, 2016), EuroSys '16, ACM, pp.~26:1--26:17.

\bibitem{widedeep16}
{\sc Cheng, H.-T., Koc, L., Harmsen, J., Shaked, T., Chandra, T., Aradhye, H.,
  Anderson, G., Corrado, G., Chai, W., Ispir, M., Anil, R., Haque, Z., Hong,
  L., Jain, V., Liu, X., and Shah, H.}
\newblock Wide \& deep learning for recommender systems.
\newblock {\em arXiv:1606.07792\/} (2016).

\bibitem{projadam14}
{\sc Chilimbi, T., Suzue, Y., Apacible, J., and Kalyanaraman, K.}
\newblock {Project Adam}: Building an efficient and scalable deep learning
  training system.
\newblock In {\em 11th USENIX Symposium on Operating Systems Design and
  Implementation\/} (2014), {OSDI}'14, USENIX.

\bibitem{grucho14}
{\sc Cho, K., van Merrienboer, B., G{\"{u}}l{\c{c}}ehre, {\c{C}}., Bougares,
  F., Schwenk, H., and Bengio, Y.}
\newblock Learning phrase representations using {RNN} encoder-decoder for
  statistical machine translation.
\newblock {\em CoRR abs/1406.1078\/} (2014).

\bibitem{torch11}
{\sc Collobert, R., Kavukcuoglu, K., and Farabet, C.}
\newblock Torch7: A matlab-like environment for machine learning.
\newblock In {\em BigLearn, NIPS Workshop\/} (2011).

\bibitem{disbelief12}
{\sc Dean, J., Corrado, G., Monga, R., Chen, K., Devin, M., Mao, M., aurelio
  Ranzato, M., Senior, A., Tucker, P., Yang, K., Le, Q.~V., and Ng, A.~Y.}
\newblock Large scale distributed deep networks.
\newblock In {\em Advances in Neural Information Processing Systems 25},
  {NIPS}'12. Curran Associates, Inc., 2012.

\bibitem{farm14}
{\sc Dragojevi{\'c}, A., Narayanan, D., Castro, M., and Hodson, O.}
\newblock {FaRM}: Fast remote memory.
\newblock In {\em 11th USENIX Symposium on Networked Systems Design and
  Implementation\/} (2014), {NSDI}'14, USENIX.

\bibitem{compro15}
{\sc Dragojevi\'{c}, A., Narayanan, D., Nightingale, E.~B., Renzelmann, M.,
  Shamis, A., Badam, A., and Castro, M.}
\newblock No compromises: Distributed transactions with consistency,
  availability, and performance.
\newblock In {\em Proceedings of the 25th Symposium on Operating Systems
  Principles\/} (New York, NY, USA, 2015), SOSP '15, ACM, pp.~54--70.

\bibitem{lstm97}
{\sc Hochreiter, S., and Schmidhuber, J.}
\newblock Long short-term memory.
\newblock {\em Neural Comput. 9}, 8 (Nov. 1997), 1735--1780.

\bibitem{herd14}
{\sc Kalia, A., Kaminsky, M., and Andersen, D.~G.}
\newblock Using rdma efficiently for key-value services.
\newblock In {\em Proceedings of the 2014 ACM Conference on SIGCOMM\/} (New
  York, NY, USA, 2014), SIGCOMM '14, ACM, pp.~295--306.

\bibitem{rdma16}
{\sc Kalia, A., Kaminsky, M., and Andersen, D.~G.}
\newblock {Design Guidelines for High Performance RDMA Systems}.
\newblock In {\em 2016 USENIX Annual Technical Conference (USENIX ATC 16)\/}
  (Denver, CO, 2016), USENIX Association, pp.~437--450.

\bibitem{fasst16}
{\sc Kalia, A., Kaminsky, M., and Andersen, D.~G.}
\newblock {FaSST: Fast, Scalable and Simple Distributed Transactions with
  Two-Sided (RDMA) Datagram RPCs}.
\newblock In {\em 12th USENIX Symposium on Operating Systems Design and
  Implementation (OSDI 16)\/} (GA, 2016), USENIX Association, pp.~185--201.

\bibitem{gpunet14}
{\sc Kim, S., Huh, S., Zhang, X., Hu, Y., Wated, A., Witchel, E., and
  Silberstein, M.}
\newblock Gpunet: Networking abstractions for gpu programs.
\newblock In {\em 11th USENIX Symposium on Operating Systems Design and
  Implementation (OSDI 14)\/} (Broomfield, CO, 2014), USENIX Association,
  pp.~201--216.

\bibitem{cifar_data}
{\sc Krizhevsky, A.}
\newblock Learning multiple layers of features from tiny images.
\newblock Tech. rep., 2009.

\bibitem{alexnet12nips}
{\sc Krizhevsky, A., Sutskever, I., and Hinton, G.~E.}
\newblock Imagenet classification with deep convolutional neural networks.
\newblock In {\em Advances in Neural Information Processing Systems 25},
  F.~Pereira, C.~J.~C. Burges, L.~Bottou, and K.~Q. Weinberger, Eds. Curran
  Associates, Inc., 2012, pp.~1097--1105.

\bibitem{ParameterServerLi2014}
{\sc Li, M., Andersen, D.~G., Park, J.~W., Smola, A.~J., Ahmed, A., Josifovski,
  V., Long, J., Shekita, E.~J., and Su, B.-Y.}
\newblock Scaling distributed machine learning with the parameter server.
\newblock In {\em 11th USENIX Symposium on Operating Systems Design and
  Implementation\/} (2014), {OSDI}'14, USENIX.

\bibitem{Jinyang13}
{\sc Mitchell, C., Geng, Y., and Li, J.}
\newblock Using one-sided {RDMA} reads to build a fast, {CPU-efficient}
  key-value store.
\newblock In {\em Proceedings of the 2013 USENIX Conference on Annual Technical
  Conference\/} (2013), {USENIX ATC}'13, USENIX.

\bibitem{Grappa15}
{\sc Nelson, J., Holt, B., Myers, B., Briggs, P., Ceze, L., Kahan, S., and
  Oskin, M.}
\newblock Latency-tolerant software distributed shared memory.
\newblock In {\em 2015 USENIX Annual Technical Conference\/} (2015), {USENIX
  ATC}'15, USENIX.

\bibitem{vggnet14}
{\sc Simonyan, K., and Zisserman, A.}
\newblock Very deep convolutional networks for large-scale image recognition.
\newblock {\em CoRR abs/1409.1556\/} (2014).

\bibitem{seq2seq14}
{\sc Sutskever, I., Vinyals, O., and Le, Q.~V.}
\newblock Sequence to sequence learning with neural networks.
\newblock In {\em Proceedings of the 27th International Conference on Neural
  Information Processing Systems\/} (Cambridge, MA, USA, 2014), NIPS'14, MIT
  Press, pp.~3104--3112.

\bibitem{inception15}
{\sc Szegedy, C., Vanhoucke, V., Ioffe, S., Shlens, J., and Wojna, Z.}
\newblock Rethinking the inception architecture for computer vision.
\newblock {\em CoRR abs/1512.00567\/} (2015).

\bibitem{nessie15}
{\sc Szepesi, T., Cassell, B., Wong, B., Brecht, T., and Liu, X.}
\newblock Nessie: A decoupled, client-driven, key-value store using rdma.
\newblock {\em Technical Report CS-2015-09, University of Waterloo, David R.
  Cheriton School of Computer Science.\/} (2015).

\bibitem{theano16}
{\sc {Theano Development Team}}.
\newblock {Theano: A {Python} framework for fast computation of mathematical
  expressions}.
\newblock {\em arXiv e-prints abs/1605.02688\/} (May 2016).

\bibitem{storm14}
{\sc Toshniwal, A., Taneja, S., Shukla, A., Ramasamy, K., Patel, J.~M.,
  Kulkarni, S., Jackson, J., Gade, K., Fu, M., Donham, J., Bhagat, N., Mittal,
  S., and Ryaboy, D.}
\newblock {Storm@Twitter}.
\newblock In {\em Proceedings of the 2014 ACM SIGMOD International Conference
  on Management of Data\/} (New York, NY, USA, 2014), SIGMOD '14, ACM,
  pp.~147--156.

\bibitem{rdmahtm15}
{\sc Wei, X., Shi, J., Chen, Y., Chen, R., and Chen, H.}
\newblock {Fast In-memory Transaction Processing Using RDMA and HTM}.
\newblock In {\em Proceedings of the 25th Symposium on Operating Systems
  Principles\/} (New York, NY, USA, 2015), SOSP '15, ACM, pp.~87--104.

\bibitem{hadoop09}
{\sc White, T.}
\newblock {\em Hadoop: The Definitive Guide}, 1st~ed.
\newblock O'Reilly Media, Inc., 2009.

\bibitem{Gram15}
{\sc Wu, M., Yang, F., Xue, J., Xiao, W., Miao, Y., Wei, L., Lin, H., Dai, Y.,
  and Zhou, L.}
\newblock {GraM}: Scaling graph computation to the trillions.
\newblock In {\em Proceedings of the Sixth ACM Symposium on Cloud Computing\/}
  (2015), {SoCC}'15, ACM.

\bibitem{tux2017}
{\sc Xiao, W., Xue, J., Miao, Y., Li, Z., Chen, C., Wu, M., Li, W., and Zhou,
  L.}
\newblock {TuX2: Distributed Graph Computation for Machine Learning}.
\newblock In {\em 14th USENIX Symposium on Networked Systems Design and
  Implementation (NSDI 17)\/} (Boston, MA, 2017), USENIX Association,
  pp.~669--682.

\bibitem{petuum15}
{\sc Xing, E.~P., Ho, Q., Dai, W., Kim, J.-K., Wei, J., Lee, S., Zheng, X.,
  Xie, P., Kumar, A., and Yu, Y.}
\newblock {Petuum}: A new platform for distributed machine learning on big
  data.
\newblock In {\em Proceedings of the 21th ACM SIGKDD International Conference
  on Knowledge Discovery and Data Mining\/} (2015), {KDD}'15, ACM.

\bibitem{cntk14}
{\sc Yu, D., Eversole, A., Seltzer, M., Yao, K., Kuchaiev, O., Zhang, Y.,
  Seide, F., Huang, Z., Guenter, B., Wang, H., Droppo, J., Zweig, G., Rossbach,
  C., Gao, J., Stolcke, A., Currey, J., Slaney, M., Chen, G., Agarwal, A.,
  Basoglu, C., Padmilac, M., Kamenev, A., Ivanov, V., Cypher, S.,
  Parthasarathi, H., Mitra, B., Peng, B., and Huang, X.}
\newblock An introduction to computational networks and the computational
  network toolkit.
\newblock Tech. rep., October 2014.

\end{thebibliography}
